\documentclass{emulateapj}

\newcommand\iso[2]{$^{\rm #1}$#2}

\def\cs22964{\mbox{CS~22964-161}}

\def\kmsec{\mbox{km~s$^{\rm -1}$}}

\def\rpro{\mbox{$r$-process}}
\def\spro{\mbox{$s$-process}}
\def\ncap{\mbox{$n$-capture}}

\shorttitle{Neutron-Capture Isotopic Fractions} 
\shortauthors{Roederer et al.}

\begin{document}

\title{Europium, Samarium, and Neodymium Isotopic Fractions 
in Metal-Poor Stars}

\author{
Ian U.\ Roederer\altaffilmark{1},
James E.\ Lawler\altaffilmark{2},
Christopher Sneden\altaffilmark{1},
John J.\ Cowan\altaffilmark{3},
Jennifer S.\ Sobeck\altaffilmark{1,4},
Catherine A.\ Pilachowski\altaffilmark{5}
}

\altaffiltext{1}{Department of Astronomy, University of Texas at Austin,
1 University Station, C1400, Austin, TX 78712-0259; 
iur,chris@astro.as.utexas.edu}

\altaffiltext{2}{Department of Physics, University of Wisconsin,
1150 University Avenue, Madison, WI 53706; jelawler@wisc.edu}

\altaffiltext{3}{Department of Physics and Astronomy, University of Oklahoma, 
Room 131, Nielsen Hall, Norman, OK 73019; cowan@nhn.ou.edu}

\altaffiltext{4}{European Southern Observatory, Karl-Schwarzschild-Strasse 2, 
D-85748 Garching bei M\"{u}nchen, Germany; jsobeck@eso.org}

\altaffiltext{5}{Department of Astronomy, Indiana University, 
Swain West 319, 727 East Third Street, Bloomington, IN 47405-7105;
catyp@astro.indiana.edu}

\begin{abstract}

We have derived isotopic fractions of europium, 
samarium, and neodymium in two metal-poor giants 
with differing neutron-capture nucleosynthetic histories.  
These isotopic fractions were measured from
new high resolution ($R\sim120,000$), 
high signal-to-noise (S/N$\,\sim\,$160--1000)
spectra obtained with the 2dCoud\'{e} spectrograph 
of McDonald Observatory's 2.7\,m Smith telescope.  
Synthetic spectra were generated using recent
high-precision laboratory measurements of hyperfine and isotopic
subcomponents of several transitions of these elements
and matched quantitatively to the observed spectra.  
We interpret our isotopic fractions by the nucleosynthesis
predictions of the stellar model, which reproduces \spro\ nucleosynthesis
from the physical conditions expected in low-mass, thermally-pulsing
stars on the AGB, and the classical method, which approximates 
\spro\ nucleosynthesis by a steady neutron flux impinging upon
Fe-peak seed nuclei.
Our Eu isotopic fraction in HD~175305 is consistent with
an \rpro\ origin by the classical method and is 
consistent with either an $r$- or an \spro\ origin
by the stellar model.
Our Sm isotopic fraction in HD~175305 suggests
a predominantly \rpro\ origin, and our Sm isotopic fraction in 
HD~196944 is consistent with an \spro\ origin.
The Nd isotopic fractions, while consistent with either \rpro\ or 
\spro\ origins, have very little ability to distinguish between 
\textit{any} physical values for the isotopic fraction in either star.
This study for the first time extends the \ncap\
origin of multiple rare earths in metal-poor stars 
from elemental abundances to the isotopic level,
strengthening the \rpro\ interpretation for 
HD~175305 and the \spro\ interpretation for HD~196944.

\end{abstract}


\section{Introduction}
\label{introduction}

A wealth of observational studies over the last decade have focused
on measuring precise abundances of neutron ($n$)-capture elements
in metal-poor stars.
The heart of this enterprise is identifying the origin of the nuclides 
with $Z\,>\,30$ in our Galaxy.
These nuclides are produced in stars through either the rapid 
($r$)-process or the slow ($s$)-process.
Rapid-process nucleosynthesis occurs in environments with high
neutron fluxes and densities, allowing many neutrons to be captured by 
existing nuclei much more rapidly than the timescales for $\beta$-decay.
Slow-process nucleosynthesis occurs where the neutron densities are lower,
with $\beta$-decay timescales generally shorter than the times
between \ncap\ events.
To build heavy nuclei through the \spro, a chain of stable or
long-lived nuclei must exist from the existing seed nucleus to the final 
product.  
The \rpro\ and \spro\ lead to the creation of different sets of heavy nuclei, 
some of which can only be created in one process or the other 
(``pure-$s$-'' or ``pure-$r$-nuclei'') and some of which are created by 
both processes \citep[see, e.g.,][]{cowan06b}.  
While the precise astrophysical site(s) of the \rpro\ have yet to be 
identified, \spro\ nucleosynthesis takes place in low- and 
intermediate-mass stars on the asymptotic giant branch (AGB)
\citep[e.g.,][]{busso99}.

\begin{figure}
\epsscale{1.0}
\plotone{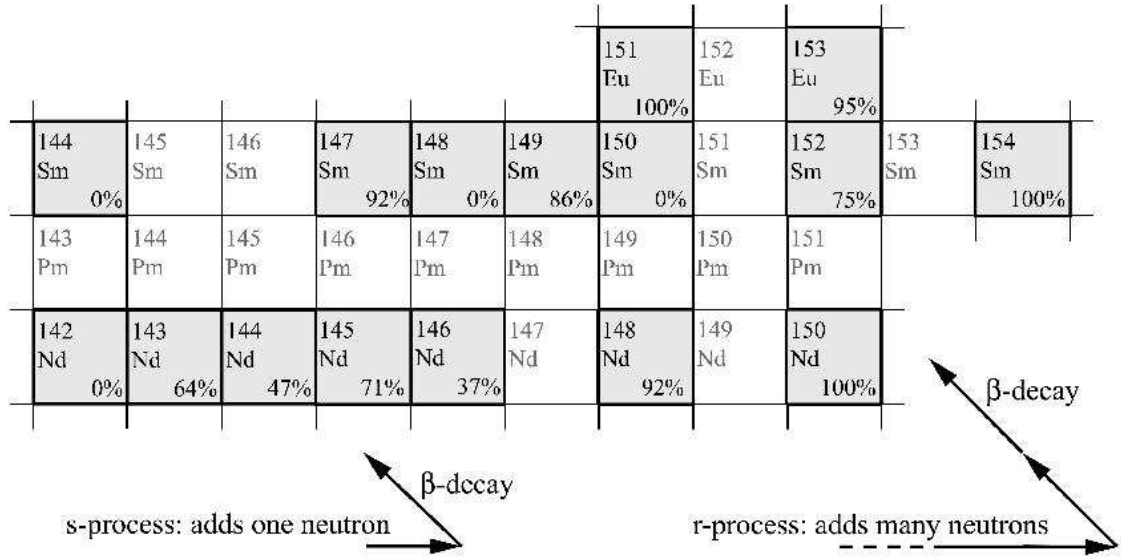}
\caption{
\label{nuclides}
A small region of the table of nuclides, showing
increasing mass number ($A$) on the horizontal axis
and increasing atomic number ($Z$) on the vertical axis.
Naturally-occurring isotopes are represented by the
gray squares.  
The figure highlights the naturally-occurring isotopes of
neodymium (Nd, $Z=60$), 
samarium (Sm, $Z=62$), and 
europium (Eu, $Z=63$);
promethium (Pm, $Z=61$) has no naturally-occurring isotopes.
We also show the predicted percentage of \rpro\ material 
in the S.~S.\ abundance for each isotope as given by
\citet{cowan06a}.
This figure illustrates the pathways by which
the $s$- and $r$-processes operate and the 
resulting differences in the $s$ or $r$ fraction
of S.~S.\ material.
}
\end{figure}

To illustrate the effects of both \ncap\ processes, in 
Figure~\ref{nuclides} we show a table of nuclides covering portions 
of the atomic number domain 60~$\leq$~$Z$~$\leq$~63. 
The \spro\ can add only one neutron before a $\beta$-decay occurs for
an unstable nucleus.
On the other hand, the \rpro\ quickly floods a target nucleus with many 
neutrons, overwhelming the $\beta$-decay rates.
When neutrons are no longer being added, the nucleus will $\beta$-decay 
repeatedly until it reaches stability.
For isotopes produced primarily in the \spro\ (e.g., \iso{146}{Nd}, 
\iso{148,150}{Sm}) a stable isotope with one less nuclide always exists.
For isotopes produced primarily or wholly by the \rpro\ 
(e.g., \iso{150}{Nd}, \iso{147,149,152,154}{Sm}), 
there is a clear path for $\beta$-decay from unstable nuclei to the 
lower right (in the table of nuclides) of the stable \rpro\ isotope, 
sometimes unreachable by the \spro.

The nucleosynthetic signatures of the $s$- and $r$-processes
stand in sharp contrast to one-another.
The Solar System (S.~S.) isotopic abundances are well-established from 
studies of CI meteorites, as summarized in, e.g.,
\citet{anders89} and \citet{lodders03}.
Two primary methods are used to decompose the S.~S.\ isotopic abundances 
into their constituent $s$- and \rpro\ origins.
In the ``standard'' or ``classical'' method, first articulated by
\citet{clayton61} and \citet{seeger65}
and with abundances later updated by, e.g., 
\citet{kappeler89,burris00,simmerer04}, and \citet{cowan06a}, 
the \spro\ is modeled as a slowly-varying function of 
neutron exposure.
This method attempts to smoothly fit the ``$\sigma N_{s}$'' curve 
(the product of the \ncap\ cross-section and the S.~S.\ $s$-process 
abundances) using lab measurements of $\sigma$ to determine $N_{s}$.
In the ``stellar'' model of \citet{arlandini99}, isotopic abundances 
for \spro\ nucleosynthesis in 1.5$-$3.0\,$M_{\sun}$ stars are 
predicted based upon nuclear physics cross sections and 
stellar model calculations for thermally-pulsing (TP) stars on the AGB.
In both of these cases, the \rpro\ abundances are then derived as residuals 
between the total S.~S.\ abundances and the $s$-process products.
Both approaches have strengths and weaknesses. 
The classical method is model-independent, but it assumes an empirical, 
smoothly-varying relationship for the $\sigma N_{s}$ abundance curve 
(which relies on the measured S.~S.\ abundances) and ignores details of 
the nuclear physics.
The stellar model does not depend on knowledge of the S.~S.\ abundance 
distribution, but is heavily dependent on nuclear physics
laboratory measurements and complex stellar AGB model calculations.

These models can also be used to predict the relative 
amounts of $s$- or \rpro\ nucleosynthesis products in metal-poor stars.
Several key rare earth elements are commonly 
used to assess $r$-/\spro\ abundance dominance in stars.\footnote{ 
The \spro\ percentages for some of the commonly-studied
rare earth elements, predicted by the stellar model 
and classical method, respectively, are:
Ba (five naturally-occurring isotopes), 81.1\% or 85.3\%;
La (one isotope), 62.1\% or 75.4\%;
Nd (seven isotopes), 58.8\% or 57.9\%;
Sm (seven isotopes), 30.7\% or 33.1\%;
Eu (two isotopes), 5.8\% or 2.3\%. }
Eu is commonly used as a surrogate for \rpro\ nucleosynthesis
because it is so predominantly composed of \rpro\ material and has 
several easily-observed spectral lines.
Similarly, Ba or La are commonly used as surrogates 
of \spro\ nucleosynthesis.
Thus, many studies have employed [Ba/Eu]\footnote{
We adopt the usual spectroscopic notations that
[A/B]~$\equiv$ log$_{\rm 10}$(N$_{\rm A}$/N$_{\rm B}$)$_{\star}$~--
log$_{\rm 10}$(N$_{\rm A}$/N$_{\rm B}$)$_{\sun}$, and
that log~$\varepsilon$(A)~$\equiv$ 
log$_{\rm 10}$(N$_{\rm A}$/N$_{\rm H}$)~$+$~12.00,
for elements A and B.
}
ratios in metal-poor stars to estimate the relative 
contributions from the $s$- and $r$-processes
\citep[e.g.,][]{spite78,gratton94,burris00,mashonkina03,
honda04b,barklem05,francois07}, which in turn
constrain models of chemical evolution in the early Galaxy 
\citep{travaglio99,travaglio01,cescutti06,kratz07}.
Recently, [La/Eu] ratios have been employed as an alternative to 
[Ba/Eu] to avoid difficulties in Ba abundance determinations
\citep[see, e.g.,][]{simmerer04,jonsell06,winckler06}.

Elemental abundances, the sums of abundances of their naturally-occurring 
constituent isotopes, are more easily measured in stars than the isotopic 
abundances; however,
isotopic abundances should be more fundamental indicators of \ncap\ 
nucleosynthesis because they can directly confront \rpro\ and \spro\ 
predictions without the smearing effect of multiple isotopes.
Much recent evidence (e.g., the growing number
of $r$-process-enhanced stars that conform to the 
scaled-S.~S.\ \rpro\ elemental abundance distribution)
supports the hypothesis of a universal \rpro\ mechanism
for elements with $Z\geq56$.
It is important to demonstrate that the \rpro\ elemental abundance 
pattern extends to the isotopic level.
Such agreement would greatly strengthen the argument for a
universal \rpro\ mechanism for the heavy \ncap\ elements.

Most isotope fractions, unlike elemental abundances, are very insensitive 
to the model atmosphere parameters.  
The wavelength of a spectral line is split by two effects,
hyperfine structure (HFS) and isotope shifts, and 
the isotopic fractions can be measured by detailed comparisons
of an observed absorption line profile to synthetic spectra
of these line substructures.
If this splitting is comparable to or greater than the 
combined effects of stellar thermal, macroturbulent, 
and rotational broadening
and the broadening caused by the spectrograph slit, 
then it may be possible to measure the isotopic fractions.
In principle, unlike with elemental abundances, 
any \ncap\ element with multiple 
naturally-occurring isotopes that are produced in different 
amounts by the $s$- and $r$-processes can be used to 
assess the relative $s$- and \rpro\ contributions to 
the stellar composition.
In practice, the availability of quality atomic data 
has limited studies of isotopic fractions of rare earth elements
in stars, but the relatively recent increase in laboratory 
spectral line substructure studies has led to the
identification of a few lines of a few elements that 
might be analyzed at the isotopic level.

\citet{cowley89} and \citet{magain93a} first suggested that the width
of the Ba~\textsc{ii} 4554\AA\ line in stellar spectra
would be sensitive to the isotopic fraction.
\citet{magain93a}, \citet{magain95}, and \citet{lambert02} all  
examined the Ba isotopic fraction in the metal-poor subgiant HD~140283.
Their measured isotopic fractions had uncertainties too large to
unambiguously discern the Ba nucleosynthetic origin in this star.
A recent series of papers by 
\citet{mashonkina99,mashonkina03} and \citet{mashonkina06}
have measured Ba isotopic fractions in thin and thick disk stars.
Their approach is to set the Ba elemental abundance from lines not affected
by HFS or isotope shifts and then to derive the isotopic fraction from the 
4554\AA\ line, whose shape is sensitive to both its elemental abundance 
and isotopic composition of Ba.
\citet{mashonkina06} clearly show that 
a higher fraction of \rpro\ Ba isotopes
is present in thick disk stars than in S.~S.\
material and thin disk stars, and they also show that
a smooth trend of increasing \spro\
contributions occurs with increasing metallicity.
Since only the 4554\AA\ line is useful for examining the 
Ba isotopic mix, and the profile of this line is known to suffer 
from non-LTE effects in metal-poor stars 
\citep[e.g.,][]{mashonkina99,short06}, we do not 
attempt to assess the Ba isotopic fractions in this study. 
The Mashonkina et al.\ studies of the Ba isotopic fraction do, however,
employ non-LTE calculations.

Eu isotopic fractions in metal-poor stars were first reported
by \citet{sneden02}, who found that the isotopic ratio in three
\rpro-rich giants was consistent with the S.~S.\ isotopic ratio.
Subsequent studies by \citet{aoki03a,aoki03b}
found that Eu isotopic ratios could marginally distinguish $s$- and \rpro\
contributions in stars that had previously
been shown to exhibit such chemical signatures.
\citet{lundqvist07} have investigated the Sm isotopic fractions 
in a very metal-poor \rpro-enriched star, qualitatively showing 
that the isotopic mix is in agreement with an \rpro\ origin.
No measurements of the Nd isotopic fraction have been made 
outside of the S.~S., and no study has yet attempted to measure 
isotopic fractions of multiple \ncap\ elements in the same star.

In our study, we analyze the isotopic
fractions of Eu, Sm, and Nd in two metal-poor giant stars.
Our observations and methods of analysis
are described in \S~\ref{observations}
and \ref{analysis}, respectively.
We summarize the available atomic data
for each species in \S~\ref{atomicdata}.
Our measurements of the elemental abundances and isotopic
fractions are presented in \S~\ref{elabund} and \S~\ref{isofrac}.
In \S~\ref{interpret} we discuss
the implications of our measurements
in the context of using isotopic fractions
of multiple species to constrain the $s$- and \rpro\
nucleosynthetic histories.
In \S~\ref{eupb} we use Eu and Pb abundances 
collected from the literature to assess the 
\spro\ model predictions for the $^{151}$Eu isotope,
since pure-$r$- and pure-$s$-process nucleosynthesis reactions
produce such contrasting abundances of these elements.
We also make suggestions for how future studies 
might utilize measurements of \ncap\ isotopic fractions
to make similar comparisons for other isotopes in \S~\ref{future}.
We summarize our results in \S~\ref{conclusions}.
In the appendices, we include comments
describing the nuances associated with
analyzing each line of each species 
(Appendix~\ref{comments}), and we include our computations
of the hyperfine structure subcomponents
for both Sm (Appendix~\ref{smappendix})
and Nd (Appendix~\ref{ndappendix}).

\section{Observations}
\label{observations}

\subsection{Selection of Target Stars}
\label{targets}

We choose two bright, metal-poor giants
for our analysis, HD~175305 and HD~196944.
The technical requirements of achieving high
signal-to-noise (S/N) ratios at very high
resolution on a telescope with a moderate aperture
mandate that we select bright ($V\,\lesssim\,$8.5) targets.
We select metal-poor giant stars for this analysis because
(1) metal-poor stars present significantly less crowding 
of strong absorption features
in the blue spectral region where the
lines of interest are concentrated,
(2) absorption lines of all species will exhibit
greater line depth in giants than in dwarfs
because of the decrease in the continuous opacity
with decreasing electron pressure in giants, and
(3) at a given metallicity, molecular formation
will be less in giants than in dwarfs.
To maximize the expected contrast in the isotopic fractions,
we have deliberately selected stars with different
chemical enrichment histories.
Although these stars have been well-studied,
no previous attempts have been made to 
examine their isotopic fractions.

HD~175305 is a metal-poor
([Fe/H]~$=-1.6$), bright ($V$~$\sim\,7.2$)
giant that was selected because it 
shows evidence of $r$-process enrichment
([Eu/Fe]~$\sim\,+0.5$, [Ba/Eu]~$<\,0.0$, [La/Eu]~$\sim\,-0.4$;
\citealt{burris00,fulbright00,cowan05}).
HD~175305 is not enriched in carbon 
([C/Fe]~$=+0.2$; \citealt{wallerstein79}).
Although only a handful of heavy $n$-capture 
abundances have been measured previously, the high
[Eu/Fe] and low [Ba/Eu] ratios are suggestive of
an \rpro\ origin.
Combining \citet{cowan05}'s
abundance measurements of osmium (Os, $Z\,=\,76$), 
iridium (Ir, $Z\,=\,77$), and platinum (Pt, $Z\,=\,78$) 
from UV spectra and
\citet{lawler07}'s measurement of hafnium (Hf, $Z\,=\,72$)
with additional
heavy \ncap\ elemental abundances measured
in our spectra will produce a more thorough
understanding of this star's enrichment history
to complement our isotopic fraction measurements.

HD~196944 is a very metal-poor
([Fe/H]~$=-2.4$), bright ($V$~$\sim\,8.4$)
giant with clear \spro\ signatures.
This star has high C, N, and O abundances
([C/Fe]~$\sim\,+1.3$, [N/Fe]~$\sim\,+1.3$, [O/Fe]~$\sim\,+1.4$;
\citealp{zacs98,aoki02,jonsell05}),
high Ba and La abundances 
([Ba/Fe]~$\sim\,+1.1$, [La/Fe]~$\sim\,+0.9$),
a low Eu abundance 
([Eu/Fe]~$\sim\,+0.2$),
and a large over-abundance of lead
([Pb/Fe]~$\sim\,+2.0$; 
\citealp{vaneck01,aoki02,vaneck03}).
HD~196944 also exhibits radial velocity variations
\citep{lucatello05}, indicating that it is a member of a
binary or multiple star system.
These characteristics have led to the
classification of HD~196944 as a 
carbon-enhanced metal-poor $s$-enriched
(CEMP-$s$) star, as defined in
\citet{beers05}.
\citet{gallino05} have shown that the abundance pattern of HD~196944 
can be reproduced by assuming mass transfer of material enriched
in \spro\ elements from a companion
star with an initial mass of 1.5\,$M_{\sun}$ that passed through the 
TP-AGB phase.

We also observed HD~122563 in our study;
at $V$~$\sim\,6.2$, this star is the brightest
very metal-poor ([Fe/H]~$=-2.8$) star in the sky.
Unfortunately the lines of Eu, Sm, and Nd proved to be too weak
to provide any assessment of the isotopic fractions.
This is unfortunate because 
\citet{honda07}, following studies of this star by
\citet{cowan05}, \citet{aoki05}, and \citet{honda06}, 
have suggested that HD~122563 and HD~88609 may be the first two known 
representatives of a class of stars 
whose \ncap\ enrichment has been dominated by the
weak \rpro, which produces a distinctly
different abundance pattern than the 
main \rpro.\footnote{Please see, e.g., 
\citealt{wasserburg96}, \citealt{qian00},
\citealt{wasserburg00}, \citealt{johnson02b},
\citealt{wanajo06}, and \citealt{kratz07} for further
discussion of these two processes.}
Our analysis of HD~122563 did allow us to assess the 
location and strength of blending features for 
some of our lines of interest.
It is possible that the Eu~\textsc{ii} lines
at 3819, 3907, and 4129\AA,
which were not covered in our spectrum, 
may give some hint to the Eu isotopic fraction.

\subsection{Description of Observations}
\label{moreobs}

For a set of lines of a given species with the same ionization state 
in a stellar atmosphere,
the relative strength of each line depends on its 
transition probability, excitation potential, wavelength, and 
the continuous opacity of the atmosphere at that wavelength
\citep[see, e.g.,][p.~389]{gray05}.
In metal-poor stars the continuous opacity 
(from H$^{-}$)
increases slowly over the visible wavelength region
as the wavelength increases;
these two terms take opposite signs in calculations of 
the line strength and roughly offset each other.
To first approximation then, the relative strengths of 
individual absorption lines are determined
by the transition properties of the lines themselves.
Figure~10 of \citet{lawler06} illustrates this point for
the case of Sm~\textsc{ii} lines.
For the \ncap\ species considered in our study,
the first-ionized state is dominant in our
stars, and most of the transitions have similar
excitation potentials ($\sim$~0.0--0.7\,eV), so 
the oscillator strengths most significantly 
affect the relative line strength.
The log($gf$) values generally decrease when moving from the near-UV 
to the near-IR spectral regions, so the strongest lines 
in stellar spectra tend to lie in the near-UV and blue
end of the visible spectrum.
This is unfortunately where the most severe
blending with other spectral features is likely to occur.
The isotope shifts increase with increasing
wavelength;
this broadens the stellar line profiles,
making it easier to measure the isotopic fraction.
The decrease in line strength, decrease in blending features,
and the increase in isotopic shifts at longer wavelengths
compete against each other, producing a range in the visible
spectrum from $\sim$~4400$-$5100\AA\ where 
isotopic fractions are most easily measured.

We acquired new observations of our 
target stars using the 2dCoud\'{e} 
cross-dispersed echelle spectrograph 
\citep{tull95} on the 2.7~m Harlan J.\ Smith Telescope
at the W.~J.\ McDonald Observatory.  
Observations of HD~122563 and HD~196944 were made on
2006 June 12--18, and observations of HD~175305
were made on 2006 August 08--12.
Sets of 30-minute exposures were taken with 
a Tektronix $2048\times2048$ CCD (``TK3'').
Around 5000\AA, to sample the full width of a line profile
($\sim$\,0.2\AA) with at least 10 points
requires $\Delta\lambda\,\approx\,$0.02 per pixel
and a two-pixel resolution of $R\,\sim\,$125,000.
This requirement dictates the spectral 
resolution necessary to complete our analysis.
Our data have FWHM resolving powers
$R\,\sim\,120,000$.
Lines in our Th--Ar comparison spectra 
can be fit with a pure Gaussian function,
and no variation in the instrumental profile
was found from night-to-night
over the course of either observing run.
Due to the limited wavelength coverage at this
setup, we have observed
each star with three grating settings to increase
the wavelength coverage.
Our spectra cover $4200 \leq \lambda \leq 6640$\AA\
in HD~122563 and HD~196944 and
$4120 \leq \lambda \leq 5890$\AA\ in HD~175305;
however, even when using multiple grating settings
our wavelength coverage is still incomplete within 
these ranges.

Image processing, order extraction, and
wavelength calibration were performed
using standard IRAF\footnote{
IRAF is distributed by the National Optical Astronomy 
Observatories, which are operated by the Association of 
Universities for Research in Astronomy, Inc., under 
cooperative agreement with the National Science Foundation.} 
tasks in the \textit{echelle} and \textit{onedspec} packages.
Echelle orders were 
cross-correlated and co-added, and 
the continuum normalization was performed with
the SPECTRE software package 
\citep{fitzpatrick87}.
Our final S/N values (per pixel) range from
$\sim 160$ at 4130\AA\ to
$\sim 330$ at 5100\AA\ in HD~175305, 
$\sim 160$ at 4200\AA\ to
$\sim 400$ at 5000\AA\ in HD~196944, and
$\sim 430$ at 4200\AA\ to 
$\sim 1000$ at 5000\AA\ in HD~122563.

\section{Atomic Data}
\label{atomicdata}

\subsection{HFS Structure and Isotope Shifts}
\label{hfsis}

HFS is primarily caused by the magnetic 
interaction between an electron and the nucleus of
an atom.
This interaction occurs when the nucleus
has a magnetic dipole moment that results from
a non-zero angular momentum, 
which occurs when the nucleus has 
an odd number of protons and/or
an odd number of neutrons.
The strength of this interaction is characterized by the 
magnetic dipole ``A'' constant, which can be measured in laboratory studies.
We also include the measured 
electric quadrupole interaction ``B'' constant in our
HFS computations, but magnetic interaction is usually dominant.
In stellar spectra, HFS is manifest as 
a broadening of a single line of a particular atom 
due to the energy shifts.
Isotope shifts, on the other hand, are slight 
rearrangements of the energy levels of one isotope
of a particular species relative to another isotope
of the same species.
The most useful lines of heavy atoms---such as those considered
in this study---usually have isotope shifts dominated by the 
field shift.
This shift is from the finite volume occupied by the nucleus of 
an atom.
Proton charge is distributed within the nucleus, and not
concentrated at the origin, hence the electric field inside 
the nucleus does not have a $1/r$ dependence.
For electrons that have a significant probability 
of existing at small $r$, such as the 
6$s$ electron in the ground state configuration
of Eu~\textsc{ii},
this can have a sizable effect.
Both the reduced mass shifts and specific mass shifts of the
upper and lower levels also contribute to the total
isotope shifts of a line.
Isotope shifts for our elements have been measured in 
laboratory experiments.
In stellar spectra, the effect of %transition
isotope shifts is to broaden the absorption
line profile when more than one isotope 
of the same atom are 
present in the stellar atmosphere.

\subsection{Europium Atomic Data}
\label{euatomic}

There are two naturally-occurring isotopes of Eu:
$^{151}$Eu (47.8\% S.~S.)
and $^{153}$Eu (52.2\%).
Both Eu isotopes exhibit HFS.
Both isotopes of Eu are produced in roughly equal
amounts in the \rpro, which dominates
over the \spro\ as the origin
of the S.~S.\ Eu.
Because there are only two Eu isotopes,
it is straightforward to measure the isotopic
fraction by measuring the fraction of 
one of these isotopes relative to the 
total amount of Eu.
The fraction $f_{151}$ is defined as
\begin{equation}
f_{151}\,=\,\frac{N(^{151}\rm Eu)}{N( \rm Eu)}.
\end{equation}
In Table~\ref{tab2}, we list the fraction of
$^{151}$Eu that would be expected if all the 
Eu present in a given star were to have
originated in only the \spro\ or the
\rpro\ for both the stellar model and classical method.
The stellar model predicts
a $f^{s}_{151}$ abundance that is consistent
with two $(r+s)$-enriched stars \citep{aoki03b,aoki06},
but the classical method predicts a $f^{s}_{151}$ abundance
that is nearly 30 times smaller.
We also note that even the predictions for the pure-\spro\ 
abundance of $^{151}$Eu vary greatly within different 
sets of classical method calculations
\citep{arlandini99,cowan06a}, so this discrepancy is not
limited to only the classical method versus stellar model predictions.
Lengthy discussion of this discrepancy 
will be taken up further in \S~\ref{eupb}.

We adopt the Eu~\textsc{ii} log($gf$) values
of \citet{lawler01b}.
The hyperfine and isotopic components are taken from \citet{ivans06}.
The HFS of the three Eu~\textsc{ii} 
lines that we consider is very wide,
$\sim\,$0.2$-$0.3\,\AA,
which facilitates our ability to measure
the isotopic fraction.

\subsection{Samarium Atomic Data}
\label{smatomic}

There are seven naturally-occurring isotopes of Sm:
$^{144}$Sm (3.1\%\,S.~S.), 
$^{147}$Sm (15.0\%), 
$^{148}$Sm (11.2\%),
$^{149}$Sm (13.8\%), 
$^{150}$Sm (7.4\%), 
$^{152}$Sm (26.7\%), and 
$^{154}$Sm (22.8\%).
The $^{144}$Sm isotope is exclusively produced by the 
$p$-process; fortunately for this analysis,
this isotope only comprises 3.1\% of the 
Sm in the S.~S., so we can neglect its contribution to
the isotopic mix.
According to the stellar model and classical method,
$^{148}$Sm and $^{150}$Sm are produced 
exclusively by the \spro,
while the \rpro\ dominates the production
of the other four isotopes.

The two odd-$A$ isotopes exhibit HFS
structure in addition to the energy shifts between
all seven isotopes.
The Sm isotope shifts are $\sim\,0.05-0.10$\,\AA,
which is slightly larger than the
Sm HFS splittings, $\sim\,0.05$\,\AA,
yet both the Sm isotope shifts and HFS splittings are 
significantly smaller than the Eu HFS splittings.
The two heaviest isotopes, 
$^{152}$Sm, and $^{154}$Sm,
have large isotope shifts relative to the
isotope shifts of the lighter five isotopes in many of the
transitions we analyze here
(cf.\ Figures 1 and 2 of \citealt{lundqvist07}).  
Therefore, it makes sense to define the
isotopic fraction in terms of these two
heaviest isotopes.
The quantity $f_{152+154}$ is defined by
\begin{equation}
f_{152+154}\,=\,\frac{N(^{152} \rm Sm)+\it N \rm(^{154}{Sm})}{N(\rm Sm)}.
\end{equation}
In Table~\ref{tab2}, we also list the isotopic fraction
of Sm that would be expected if all the 
Sm present in a particular star were to have
originated in only the \spro\ or the
\rpro\ for both the stellar model and 
classical method.
Note that this definition more easily distinguishes
pure-$s$ and pure-$r$ isotopes of Sm than the $f_{\rm odd}$
definition suggested in \S~7 of \citet{lawler06}.
In addition, the pure-$s$ and pure-$r$ isotope fractions
predicted by the stellar model and classical method
do not differ significantly when expressed in
terms of $f_{152+154}$.

We have computed the HFS structure patterns for 13 
Sm lines in this wavelength range using the
hyperfine A and B constants and isotope shifts
of \citet{masterman03}, when available, 
supplemented with additional values from \citet{lundqvist07} and 
others as described in Appendix~\ref{smappendix}.
Three of these lines (4719, 5052, and 5103)
do not have complete sets
of hyperfine A and B constants reported in the
literature.  
To gauge the impact of the HFS splitting on our
stellar syntheses, we have compared two syntheses
of lines with full sets of hyperfine A and B constants,
one including the HFS splitting and isotope shifts 
and one including only the isotope shifts.
Syntheses with a lower fraction of $f_{152+154}$
are slightly more broadened in the case of
full HFS treatment than when only the isotope 
shifts are considered.
The differences are small; 
the synthesis including HFS is $\approx$\,5$-$6\%
wider at FWHM than the synthesis with no HFS
(when the percentage of isotopes exhibiting HFS
is maximized in the synthesis).
Nevertheless we recommend that lines lacking complete sets of
hyperfine A and B constants should not be used
to measure the isotopic fraction by fitting the shape
of the line profile;
however, they should still be useful for other methods
of measuring the isotopic fraction and for elemental
abundance analyses and are included in the discussion
here for this reason.
We adopt the Sm~\textsc{ii} log($gf$) values of
\citet{lawler06}.

\subsection{Neodymium Atomic Data}
\label{ndatomic}

Nd, like Sm, has seven naturally-occurring isotopes in
S.~S.\ material:
$^{142}$Nd (27.2\%\,S.~S.),
$^{143}$Nd (12.2\%),
$^{144}$Nd (23.8\%),
$^{145}$Nd (8.3\%),
$^{146}$Nd (17.2\%),
$^{148}$Nd (5.7\%), and 
$^{150}$Nd (5.6\%).
The lightest isotope is only produced in the 
\spro,
the two heaviest isotopes are almost exclusively
produced in the \rpro, and 
the other four isotopes are produced in a
combination of the two processes.  
The two odd-$A$ isotopes exhibit HFS structure,
like Sm.
\citet{rosner05} have measured the isotope shifts
and HFS A and B constants for Nd
transitions between 4175 and 4650\AA.
For the lines examined in our study,
the Nd~\textsc{ii} HFS is comparable 
in scale to the Sm~\textsc{ii} HFS, 
$\lesssim\,0.05$\AA,
while the Nd~\textsc{ii} isotope shifts are
relatively small, usually only $\lesssim\,0.025$\AA.
\citet{denhartog03} measured log($gf$) values
for many of these transitions and noted 
that several of the yellow-red lines
(mostly redward of 5100\AA) exhibited a 
doublet or triplet structure in their FTS profiles.
%This suggests that the isotope shifts of Nd, like Sm,
%generally become larger when moving from the blue to the
%red spectral regions.
%Unfortunately, like Sm, the log($gf$) values for these
%Nd transitions generally decrease over the same
%range.

We collect the Nd isotopes into fractions that 
might provide a good indication of pure-$s$- and
pure-$r$-nucleosynthesis while making sensible choices
about which sets of isotopes are shifted the most from one another.
We adopt the Nd isotopic fraction defined by 
\begin{equation}
f_{142+144}\,=\,\frac{N(^{142}\rm Nd)+ \it N \rm (^{144}Nd)}{N(\rm Nd)}.
\end{equation}
Predictions for this isotopic fraction
are displayed in Table~\ref{tab2}.
Note that the pure-$s$ and pure-$r$ predictions of the 
stellar model and classical method generally agree well
for $f_{142+144}$.
In Appendix~\ref{ndappendix} we present our calculations of
the HFS structure patterns for 6 Nd lines in the wavelength
region covered by our spectra.

\section{Analysis}
\label{analysis}

\subsection{Radial Velocities}
\label{radvel}

To measure the radial velocities,
we find the centroid of $\sim$\,17$-$19 
Fe~\textsc{i} lines and calibrate to the
wavelengths in \citet{learner88}\footnote{
\citet{whaling02} propose a change in the 
\citet{learner88} Fe~\textsc{i} wavenumber scale,
who calibrated to the absolute Ar~\textsc{ii} wavenumber scale
from \citet{norlen73}.
The proposed change, a multiplicative correction of
1$+$67(8)$\times10^{-9}$, amounts to a correction of
$\sim\,0.001-0.002$\,cm$^{-1}$ or $0.3-0.5$\,m\AA\ in the
visible spectral region examined in our study.
This correction is negligible for our purposes.
},
which are accurate to better than 0.3\,m\AA.
We find our wavelength scale zeropoint from a small number of
telluric lines in the reddest orders of our spectra.
Heliocentric velocities for each observation 
are calculated using
the IRAF \textit{rvcorrect} task.
For HD~175305 and HD~122563, we measure heliocentric radial velocities
of  $-184.2\,\pm\,0.3$\,\kmsec and $-26.0\,\pm\,0.3$\,\kmsec, 
respectively, which are both consistent with 
measurements made by previous studies to suggest no 
radial velocity variation 
(HD~175305: \citealt{carney03,nordstrom04};
HD~122563: \citealt{bond80,gratton94,barbuy03,honda04a,aoki05,aoki07}).
For HD~196944, we measure a heliocentric radial velocity
of $-166.4\,\pm\,0.3$\,\kmsec.
The scatter in our radial velocity measurements
of this star 
from individual observations over the very small date range
2006 June 12--18 is consistent with a single value.
\citet{lucatello05} compiled recent radial velocity measurements
for this star (ranging from  $-174.76\,\pm\,0.36$\,\kmsec,
\citealt{aoki02} to $-168.49\,\pm\,0.11$\,\kmsec,
\citealt[][see additional references therein]{lucatello05})
and concluded that it does exhibit 
radial velocity variations, consistent with their 
assertion that all carbon-enhanced, $s$-process-rich,
very metal-poor stars are members of binary (or multiple)
star systems.
Our measurement supports the inclusion of HD~196944 
in this category of stars.

\subsection{Equivalent Width Measurements}
\label{ewmeasure}

Our high-resolution and high-S/N observations 
of these stars are superior to spectra
analyzed in previous studies, so we 
re-derive the atmospheric parameters for these stars.
We combine the linelists and adopt 
log($gf$) values from
\citet{fulbright00}, \citet{honda04a}, and \citet{ivans06}.
Equivalent widths of isolated lines are measured 
in IRAF by fitting Gaussian profiles 
or by direct integration.
We later discard the equivalent width measurements
of lines that are approaching saturation,
as indicated by their position 
on an empirical curve-of-growth.
In each of our stars,
equivalent widths are measured for 
$\sim$\,40$-$50 Fe \textsc{i} lines,
$\sim$\,8 Fe \textsc{ii} lines,
$\sim$\,10 Ti \textsc{i} lines, and
$\sim$\,7 Ti \textsc{ii} lines.
A comparison of our equivalent width measurements
with previous high-resolution studies of these
reveals that there are no systematic differences 
in the measured equivalent widths between
our study and their studies.
We find mean offsets of only
$\Delta\,$EW$\,\equiv$\,(previous study)$-$(our 
study)$\,=\,+0.5\,\pm\,1.2$\,m\AA\ for HD~122563 \citep{honda04a},
$\Delta\,$EW$\,=\,-1.0\,\pm\,2.2$\,m\AA\ for HD~175305 \citep{fulbright00}, and
$\Delta\,$EW$\,=\,-0.8\,\pm\,2.3$\,m\AA\ for HD~196944 \citep{zacs98}.

\subsection{Atmospheric Parameters}
\label{atmosphere}

We use \citet{kurucz93} model atmospheres
without convective overshooting for 
HD~175305 and HD~196944 and with overshooting
for HD~122563 because its metallicity is beyond
the range of atmospheres provided without overshooting.
Interpolation software for the Kurucz grids has been
kindly provided by A.\ McWilliam and I.\ Ivans (2003,
private communication).

We have obtained $V-J$, $V-H$, and $V-K$ colors for our 
stars from the SIMBAD\footnote{
This research has made use of the SIMBAD database,
operated at CDS, Strasbourg, France.} and
2MASS\footnote{
This research has made use of the NASA/ IPAC Infrared 
Science Archive, which is operated by the Jet Propulsion 
Laboratory, California Institute of Technology, under 
contract with the National Aeronautics and Space Administration.}
\footnote{
This publication makes use of data products from the Two Micron 
All Sky Survey, which is a joint project of the University of 
Massachusetts and the Infrared Processing and Analysis 
Center/California Institute of Technology, funded by the 
National Aeronautics and Space Administration and the 
National Science Foundation.} \citep{skrutskie06}
databases.
We adopt zero reddening values for these stars as
recommended by previous studies
(HD~122563: \citealt{honda04b};
HD~175305: \citealt{wallerstein79};
HD~196944: \citealt{zacs98}, \citealt{jonsell05}).
We derive effective temperatures for these stars 
using the empirical color-$T_{\rm eff}$ calibrations
given by \citet{ramirez05b},
using interpolation software kindly provided by
I.\ Ram\'{i}rez\footnote{
Available online:
\url{https://webspace.utexas.edu/ir68/teff/dataII.htm}}.
Their work recalibrated and
extended the range of color and metallicity applicability of 
earlier work by \citet{alonso96b,alonso99b},
who employed the infrared flux method to determine 
effective temperatures of F, G, and K dwarfs and giants.
There is a rather small amount of scatter
among giants in the temperature and metallicity ranges
we are using 
(see \citealt{ramirez05b}'s Figure~4).
The uncertainties intrinsic to the scatter of these 
calibrations 
and in the scatter of the temperatures
predicted by each individual color 
lead us to adopt $\pm\,100$\,K 
as the uncertainty in our effective temperatures.

To determine the remainder of the model atmosphere parameters
we use the most recent version of the LTE
spectrum analysis code MOOG \citep{sneden73}.
We allow for slight ($\sim$\,20\,K) 
modifications to the photometric
temperatures to ensure that the abundances derived
from lines with both high and low excitation potentials 
agree in our final atmospheric model.
We derive the surface gravity by requiring that the 
abundances derived from neutral and ionized lines of
Fe and Ti agree;
we adopt $\pm\,0.3$ as the uncertainty in log($g$).
The microturbulence is measured by requiring that 
the abundances derived from strong and weak lines 
of each atomic species agree; 
we adopt $\pm\,0.3$\,\kmsec\
as the uncertainty in our microturbulence.
These parameters are varied iteratively 
until we arrive at our final values.

We employ synthetic spectra to 
measure the macroturbulent broadening
in our stars by fitting
the line profiles of $\sim$\,5$-$10 
clean Fe \textsc{i} and Fe \textsc{ii}
lines in each star.
Uncertainties of 0.3$-$0.4\,\kmsec\ were found
for the macroturbulent velocities.
Our line profiles are well-fit 
by the convolution of the Gaussian (instrumental) 
and macroturbulent (stellar) broadening terms.

Our metallicity is defined to be the 
derived Fe \textsc{i} abundance,
with an adopted uncertainty of $\pm\,0.15$ dex.
This uncertainty represents the systematic uncertainties
associated with determining $T_{\rm eff}$ ($\sim 0.09$--0.12~dex)
and the intrinsic line-to-line scatter ($\sim 0.10$~dex).
Each of these stars are known to be $\alpha$-enhanced
(HD~175305: [$\alpha$/Fe]~$=+0.3$, \citealt{fulbright00};
HD~196944: [$\alpha$/Fe]~$=+0.35$, \citealt{zacs98};
HD~122563: [$\alpha$/Fe]~$=+0.4$, \citealt{honda04b}).
We have increased the overall metal abundance
in our model atmosphere, [M/H], by 0.2 dex
relative to the Fe~\textsc{i} abundance to account for the
extra electron-donating $\alpha$-elements
\citep[see, e.g.][]{brown91,sneden94}\footnote{
We have performed a test to compare abundances measured from
equivalent widths of Ti, Fe, and Eu
using three different Kurucz model atmospheres.
For a typical metal-poor star, we adopt $T_{\rm eff}=5000$\,K,
$\log(g)=2.0$, and $v_{t}=2.0$\,km\,s$^{-1}$ in all three models.
For atmosphere 1, [Fe/H]~$=-2.0$ and [$\alpha$/Fe]~$=+0.0$.
For atmosphere 2, [Fe/H]~$=-1.8$ and [$\alpha$/Fe]~$=+0.0$.
For atmosphere 3, [Fe/H]~$=-2.0$ and [$\alpha$/Fe]~$=+0.4$.
Atmospheres 1 and 2 use the Kurucz grid of $\alpha$-normal atmospheres,
while atmosphere 3 uses the Kurucz grid of $\alpha$-enhanced atmospheres.
The derived abundances are equivalent to within 0.02~dex in all cases,
and this systematic effect is completely negligible for purposes of our study.
}.
For elemental abundance ratios 
(e.g., [X/Fe]) measured in our study,
we reference neutral species to the Fe~\textsc{i} 
abundance and 
ionized species to the Fe~\textsc{ii} abundance.

Our final model atmosphere parameters are listed in 
Table~\ref{tab1}.
For comparison, we show the atmospheres derived in 
previous studies of these stars.
Table~\ref{tab1} indicates that our atmospheric parameters
are in good agreement with these studies.

\subsection{Line Profile Analysis}
\label{lineanalysis}

To measure the isotopic fractions,
the observed absorption lines are fit by
synthetic spectra. 
Although our line lists are initially 
generated from the extensive lists of \citet{kurucz95}\footnote{
Available online: \url{http://kurucz.harvard.edu/linelists.html}},
we employ experimental wavelengths and
log($gf$) values for all lines of interest
and blended features whenever possible.
Our Eu, Sm, and Nd lines are synthesized
by accounting for the individual 
hyperfine and isotopic components;
the isotope fraction and elemental abundance
are varied to provide a best fit to the 
observed profiles.

At the spectral resolution employed in this study,
some of our unblended Fe~\textsc{i} and \textsc{ii}
lines exhibit a slight profile asymmetry.
This observation is consistent with our
assumption that the macroturbulent velocity
arises, at least in part, from large-scale
convective motions in the stellar photosphere.
These asymmetries account for $\approx$~1$-$2\%
of the total equivalent width of the line,
and the typical bisector amplitude is 
$\approx$~200$-$300\,m\,s$^{-1}$.
The shapes of our line bisectors are consistent
with those observed for K giants, e.g., Arcturus, by 
\citet{dravins87}.  
We have not explored any attempts to introduce 3D modeling
of the stellar atmospheres into our analysis;
such models are not yet available for metal-poor giant stars.
In this study we restrict ourselves to the
use of one-dimensional, plane-parallel
model atmospheres computed for the LTE case.

In contrast to the challenge of detecting the presence of
$^{6}$Li from the Li~\textsc{i} 6707\AA\ line profile,
which likely requires 3D-NLTE modeling to account for the 
convective asymmetry that can mimic the presence of $^{6}$Li
\citep{cayrel07}, the change in the line profile that we analyze 
is very different.
The presence of the $^{6}$Li component may account for only a few 
tenths of a percent of the total flux in the red wing of the 
$^{7}$Li line, which is a very subtle effect.
The magnitude of the HFS for Eu causes gross 
changes in the shape of the entire line profile (see \S~\ref{euiso}).
Sm and Nd each have seven naturally-occuring isotopes,
so the shape of the line profile will be asymmetric
along all parts of the line and not just along the red wing
where convective asymmetries are most prominent.
Therefore our choice to compute 1D synthetic line profiles 
in LTE will not significantly affect our results for these species.

\subsubsection{Detailed Examination of the Absorption Line Profile}
\label{linefits}

Isotope shifts and HFS alter the shape of the line profile,
and the isotopic fraction can be assessed by examining
the line profile from detailed comparisons between 
the observed and synthetic spectra.
This method has been used previously
to assess the Eu isotopic fraction by \citet{sneden02}
and \citet{aoki03a,aoki03b}.
We quantify our measurements of the isotopic fractions
using a $\chi^{2}$ algorithm 
(after, e.g., \citealt{smith01,aoki03a,asplund06}),
\begin{equation}
\chi^{2} = \left\langle
\frac{(O_{i}-C_{i})^{2}}{\sigma^{2}_{i}}\right\rangle .
\end{equation}
Here $O-C$ is the difference between the 
observed and synthetic spectra at the $i$th 
spectrum point and $\sigma$ is defined as
$\sigma_{i} = [$(S/N)$ \times (f_{i})^{1/2}]^{-1}$,
where S/N is the approximate signal-to-noise ratio of the 
continuum and $f$ is the depth of the 
$i$th point relative to the continuum.
Our best estimate of the isotopic fraction
and abundance
for each line is given by minimizing $\chi^{2}$.
Uncertainties are estimated at 1, 2, and 3\,$\sigma$
confidence intervals from 
$\Delta\chi^{2}=\chi^{2}-\chi^{2}_{min}=1$, 4, and 9, 
respectively.
We consider the number of degrees of freedom in the
fit to be the number of points in the spectral window
plus the continuum normalization, abundance, and 
wavelength offset, all of which were allowed to vary in our fits.
Reduced $\chi^{2}$ values are $\sim 1$, differing 
by less than a factor of two, which is still larger than 
would be expected for a normal distribution of errors
\citep[e.g.,][]{press92}.
We attribute the scatter in the reduced $\chi^{2}$ values
to our estimation of the S/N ratio.
The \textit{relative} $\chi^{2}$ values for a given line, 
represented by $\Delta\chi^{2}$, however,
do exhibit well-determined minima.
We fix the macroturbulence, as determined in \S~\ref{atmosphere},
but we also estimate the uncertainties 
in the isotopic fraction introduced by the uncertainty in the 
macroturbulence, $V_{\rm macro}\,\pm\,\Delta\,V_{\rm macro}$.
Only our Eu and Sm lines with complete sets of hyperfine constants
(see \S~\ref{atomicdata}) can be assessed by this method.

\subsubsection{Set the Absolute Wavelength Scale 
from Nearby Wavelength Standards}
\label{wavelengthstandard}

In principle, the absolute wavelengths of the observed and
synthetic spectra can by matched using wavelength standard
Fe~\textsc{i} and Ti~\textsc{ii} lines, and then
the peak wavelength of the rare-earth line of interest
can be measured.  
The isotopic fraction can be assessed because the 
isotope shifts---much more than the HFS---affect the
peak absorption wavelength.  
\citet{lundqvist07} have used this method to 
qualitatively demonstrate that the Sm isotopic 
mix in CS~31082-001 is consistent with an \rpro\ origin.

For our Sm lines, a small wavelength shift 
($\sim$~0.02$-$0.04\AA) 
is the dominant effect of changing the isotopic fraction; 
for our Nd lines, the shift is even smaller ($\sim$~0.01\AA).
In both cases, 
broadening of the line profile is a secondary effect.
Therefore, given the small magnitudes of these shifts,
it is critical that we match the wavelength
scales of our synthetic and observed spectra with 
great accuracy.
The absolute wavelengths of some of our Sm HFS components are
accurate to $\pm$\,1\,m\AA, as determined by the FTS measurements
of $^{154}$Sm by \citealt{lundqvist07} (see Appendix~\ref{smappendix}),
while the absolute wavelengths of the Sm lines
not included in that study are accurate to $\approx$~2\,m\AA.
Unlike Sm, however, the atomic data for Nd does not
include FTS measurements of individual isotope positions,
so our reported absolute wavelengths of Nd components
are based solely on measurements of the Nd energy levels.
To avoid mixing measurements of energy levels from different
studies, we adopt the energy levels from \citet{blaise84}.
They estimate an uncertainty of $\pm\,0.005$\,cm\,$^{-1}$
for each level.
Doubling this uncertainty translates to an uncertainty
of $\sim$~1.5--2.0\,m\AA\ across the region
where our Nd lines are located.
We note that the relative 
HFS component wavelengths---measured from
FTS spectra---are more accurate than this in all cases.
The absolute wavelengths of the wavelength standard lines
(Fe~\textsc{i}: \citealt{learner88} and \citealt{nave94}; 
Ti~\textsc{ii}: \citealt{pickering01,pickering02})
have a reported absolute accuracy in their FTS
measurements of $\sim\,\pm$\,0.001\,cm$^{-1}$
($\lesssim$\,0.3\,m\AA) in our wavelength regions.
The uncertainties from our wavelength
scale and co-addition of individual orders are
no larger than $\pm$\,0.9\,m\AA, and we have 
verified that, after the wavelength solution has been applied,
the Th-Ar emission lines are correctly identified to within
~$\sim\,\pm\,$\,1\,m\AA.
All of these uncertainties are sufficiently small to enable 
reliable measurement of the isotopic fraction.

In each echelle order containing a rare-earth line of interest,
we locate 1--6 wavelength standard lines and match the wavelength
scales of the observed and synthetic spectra at each of these lines,
typically with a precision of $\sim$~1\,m\AA\ per line.  
When multiple wavelength standards are used, the scatter 
in our measured wavelength offsets necessary to ensure the 
matching of each line is $\sim$~5\,m\AA,
much greater than anticipated.
This scatter appears to be independent of the ionization state,
species, excitation potential, 
and relative strength of the wavelength standards,
each of which could be expected to introduce small 
systematic offsets among lines with these different characteristics
(see, e.g., \citealt{dravins81,dravins86,asplund00}.
The isotopic fraction we would derive from our rare-earth
lines of interest is therefore entirely dependent
upon our choice of which wavelength standard line(s)
we adopt and how many of these lines are present
in the echelle order with the rare-earth line.

To further understand this matter, 
we have also attempted to match the observed and synthetic spectra
for wavelength standards in the Solar spectrum.  
We compare both the \citet{kurucz84} integrated-disk Solar spectrum,
obtained with the FTS on the McMath Solar Telescope at 
Kitt Peak National Observatory, and 
daylight sky spectra, obtained using the same instrumental
configuration on the McDonald Observatory 2dCoud\'{e} echelle
spectrograph as our stellar observations and reduced in an identical fashion.
It is well-known that astronomical echelle spectrographs
and cameras can introduce a variety of non-linear distortions to the 
wavelength scale, whereas
%, including 
%small changes to the ruling constants during the replication
%of the echelle grating,
%distortions from the camera optics,
%the (non-)flatness of the CCD detector, and 
%CCD pixel-to-pixel spacing variations.
the spectrum obtained with the FTS should be immune to many of
these distortions.
We synthesize a spectral region from 4997--5010\AA\ 
(one echelle order) using an interpolated
\citet{holweger74} empirical Solar model photosphere.
We apply our matching technique to the five unblended 
Fe~\textsc{i} wavelength standard lines in common to the two spectra.
We measure the same line-to-line wavelength offsets for these lines
in both spectra (agreement within our measurement precision,
$\sim$~1\,m\AA), suggesting that the wavelength scale
of our 2dCoud\'{e} spectra is consistent with the Kitt Peak 
FTS spectra.
Nevertheless, the scatter in the absolute wavelength scale
as determined from each of these lines in the FTS spectrum
is \textit{still} $\sim$~5\,m\AA, indicating
that there is a more fundamental problem that does not result
from our spectral reduction techniques or our ability to
match the observed and synthetic spectra.
Unable to further identify and quantify the source of this discrepancy,
we are unable to proceed with confidence by this method 
of analysis.

\subsubsection{Set the Relative Wavelength Scale
at the Point of Insensitivity to the Isotopic Mix}
\label{wavelengthpoint}

We also develop one additional method of quantifying the 
isotopic fraction that does not rely on nearby wavelength
standard lines or fits to the shape of the line profile.
The reason for developing such a method is our desire to assess
the isotopic mix from all lines with 
maximum isotopic shifts $\gtrsim 15$\,m\AA\ and from which an 
elemental abundance can be measured.
This allows us to examine additional Sm lines that do not have 
full sets of measured hyperfine A and B constants as well as 
all of the Nd lines.

\begin{figure}
\epsscale{1.0}
\plotone{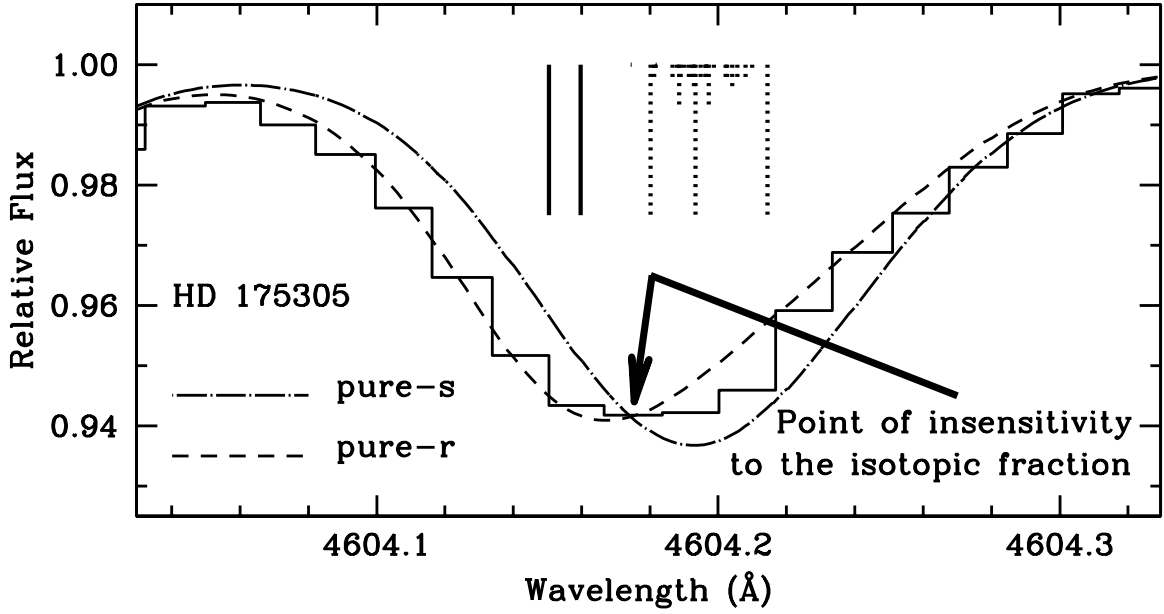}
\caption{
\label{4604shift}
One method used to match the observed and synthetic spectra, 
displaying how the point of insensitivity has been matched to the
observed spectrum.
The pure-$s$- and pure-$r$-syntheses are indicated by 
dot-dashed and dashed lines, respectively.
The observed spectrum is represented by the histogram.
The point at which the pure-$s$- and pure-$r$-process syntheses
intersect is defined to be the point of insensitivity to the 
isotopic mix.
We adjust the wavelength of the synthetic spectra so that the point 
of insensitivity coincides with the
center of the nearest pixel in the observed spectrum.
The isotopic fraction can then be measured using the 
fitting algorithm.
%The uncertainty in the derived isotopic fraction is dominated
%by the uncertainty in the matching, given by 
%$\pm$ 1/2 of a pixel width.
The vertical sticks represent the positions and 
relative strengths of the individual hyperfine components;
solid sticks represent the $^{152}$Sm and $^{154}$Sm isotopes,
while dashed sticks represent the remaining five isotopes.
}
\end{figure}

This method attempts to match the observed and synthetic
spectrum using the point in the synthetic
spectrum that is insensitive to the $s$- or \rpro\ mix.
First, for each line, an isotopic fraction is assumed in the synthesis, 
and we always adopt $f=0.50$ by default.  
Then an elemental abundance is derived from a synthesis with 
this isotopic fraction.
Using this elemental abundance, two new syntheses (one
with a pure-$s$-process mix of isotopes and one with a
pure-$r$-process mix) are created.  
The intersection of these two syntheses is defined to be
the point of insensitivity to the isotopic fraction.
The relative wavelength offset between the point of insensitivity 
and the observed spectrum is then adjusted 
until the point of insensitivity is coincident with the flux of the 
nearest pixel of the observed spectrum.
%(or the fictitious line connecting the midpoints
%of the flux of the two nearest pixels if the point of insensitivity
%does not lie at the same depth as an observed flux).
Next, we adopt this relative wavelength offset and generate 
a set of synthetic spectra with a range of isotopic fractions but
the same elemental abundance.
A new isotopic fraction is then measured by minimizing the 
sum of the squares of the differences between the observed 
and synthetic spectra.
This isotopic fraction is then used as the input to derive
an elemental abundance, and the process is repeated until the 
elemental abundance and isotopic fraction have each converged.  
We show an example where the point of insensitivity has been 
matched to the observed spectrum in Figure~\ref{4604shift}.

The uncertainty in the matching process is given by the
width of one pixel; therefore
the uncertainty in the derived isotopic fraction is 
estimated by shifting the synthetic spectrum relative to the
observed spectrum by $\pm\,w_{\rm pixel}/2$.  
Even at a spectral resolution of $R~\sim~120,000$, 
the width of one pixel is still a substantial fraction
of the width of the absorption line itself. 
This method also places a disproportionate
amount of weight on the one or two flux measurements
closest to the point of insensitivity, rather than 
taking account of the information from the full width of the line
to match the observed and synthetic spectra.
The precision achieved by this method is hardly adequate to 
clearly assess the isotopic mix---particularly
for Nd---but it does provide a consistent set of 
measurements for all of our lines of Sm and Nd.

\section{Neutron-Capture Elemental Abundances}
\label{elabund}

Our mean elemental abundance of Eu in HD~175305,
[Eu/Fe]~$=+0.46\,\pm\,0.07$, is in good agreement with 
\citet{fulbright00}'s abundance, [Eu/Fe]~$=+0.48\,\pm\,0.105$,
and \citet{cowan05}'s abundance, [Eu/Fe]~$=+0.55\,\pm\,0.10$.
Our synthesis of the Eu~\textsc{ii} 4205\AA\ line 
in HD~122563 does
permit measurement of the Eu elemental abundance,
[Eu/Fe]~$=-0.55\,\pm\,0.06$, 
assuming $^{12}$C/$^{13}$C$\,=\,5$ \citep{lambert77,honda04b} and 
a Eu isotopic fraction of $f_{151}\,=\,0.5$.
The Eu abundance changes by $\pm\,0.03$~dex
when the assumed isotopic ratio is varied by $\pm\,0.2$~dex.
This elemental abundance agrees very well with the value found by
\citet{honda06}, [Eu/Fe]~$=-0.52$,
and with many previous studies of this star.
The 4205\AA\ line in HD~196944 is far too sensitive to the
C abundance to derive a meaningful Eu elemental abundance.

We measure Sm elemental abundances of
[Sm/Fe]~$=-0.56\,\pm\,0.12$ from 1 line in HD~122563,
[Sm/Fe]~$=+0.37\,\pm\,0.11$ from 9 lines in HD~175305, and
[Sm/Fe]~$=+0.88\,\pm\,0.08$ from 6 lines in HD~196944.
Previous studies of
Sm elemental abundances in these stars have found
[Sm/Fe]~$=-0.40\,\pm\,0.17$ in HD~122563 \citep{honda06} and
[Sm/Fe]~$=+0.78\,\pm\,0.23$ in HD~196944 \citep{aoki02};
Sm has not been measured previously in HD~175305.
Unlike the uncertainties in the isotopic fractions, the
uncertainties in the elemental abundances are dominated
by the systematic uncertainties in the model atmosphere parameters
and are affected somewhat less by 
internal scatter from the log($gf$) values.

We measure Nd elemental abundances of
%[Nd/Fe]~$=$ from n lines in HD~122563,
[Nd/Fe]~$=+0.22\,\pm\,0.13$ from 12 lines in HD~175305 and
[Nd/Fe]~$=+0.95\,\pm\,0.13$ from 11 lines in HD~196944.
Previous studies of Nd elemental abundances in these
stars have found 
%[Nd/Fe]~$=-0.69\,\pm\,0.10$ \citep{honda06} in HD~122563
[Nd/Fe]~$=+0.34$ \citep{burris00} in HD~175305 and 
[Nd/Fe]~$=+0.93\,\pm\,0.09$ \citep{vaneck01},
[Nd/Fe]~$=+0.94\,\pm\,0.17$ \citep{zacs98}, and
[Nd/Fe]~$=+0.86\,\pm\,0.20$ \citep{aoki02} in HD~196944.

\begin{figure}
\epsscale{1.0}
\plotone{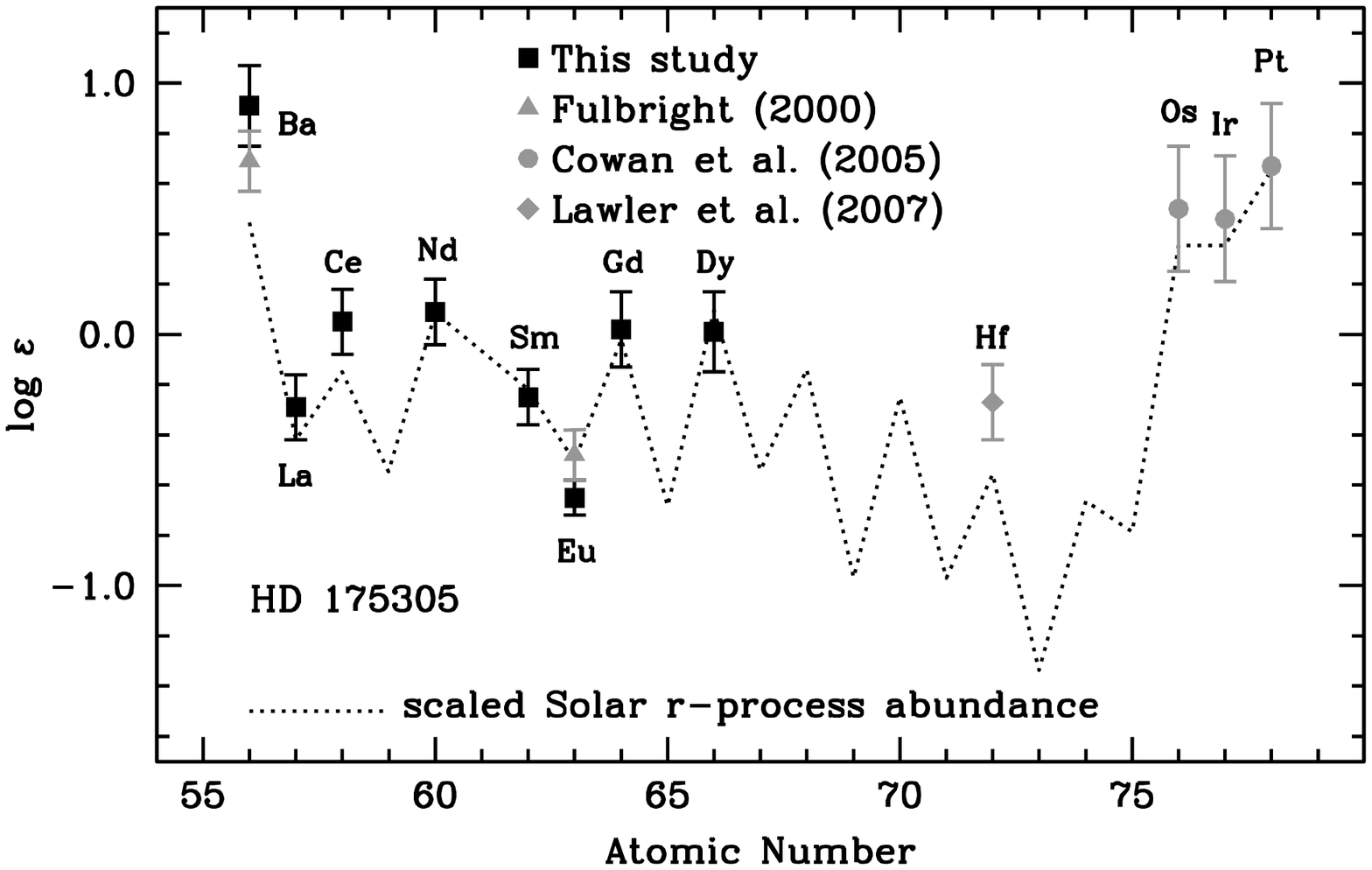}
\caption{
\label{175305ncap} 
Heavy $n$-capture elemental abundances for HD~175305.
Black squares indicate our observed abundances.
Gray triangles indicate the observed
Ba and Eu abundances from \citet{fulbright00}.
Gray circles indicate the observed
Os, Ir, and Pt abundances from \citet{cowan05}.
The gray diamond indicates the observed
Hf abundance from \citet{lawler07}.
The dotted line shows the scaled S.~S.\
\rpro\ abundances of \citet{simmerer04},
normalized to \citet{fulbright00}'s 
Eu abundance.
The figure suggests that the scaled Solar
\rpro\ abundances provide a
respectable fit to the observed abundances
beyond Ba.
}
\end{figure}

We have also measured elemental abundances for
five additional heavy \ncap\ elements in 
HD~175305.
These abundances are listed in Table~\ref{tab3}
along with our Eu, Sm, and Nd abundances.
The uncertainties listed in the table reflect the 
uncertainties in the atmospheric parameters, continuum 
placement, line-to-line scatter, and $\log(gf)$ values,
all added in quadrature.
In cases where only one line of a species was available for
analysis, we substitute the uncertainty in the spectral synthesis
fit to the observed spectrum in place of the line-to-line 
statistical scatter in this formula.
In Figure~\ref{175305ncap} we plot our
elemental abundances, combined with measurements
from \citet{fulbright00}, \citet{cowan05},
and \citet{lawler07}, 
against the scaled S.~S.\ \rpro\
abundance pattern of \citet{simmerer04}.
While not a perfect fit,
this figure suggests that the scaled S.~S.\ \rpro\ 
abundance pattern
provides a respectable fit of the
observed \ncap\ abundances beyond Ba.
HD~175305 may not be a star with pristine \rpro\ material,
but the \rpro\ appears to have contributed the majority
of the \ncap\ species present.
We have also used the Keck~I HIRES spectrum of HD~175305,
obtained for the \citet{cowan05} study, to determine 
an upper limit for Pb from the 4057\AA\ line.

\section{Neutron-Capture Isotopic Fractions}
\label{isofrac}

We emphasize that measurements of isotopic fractions
are highly insensitive to the choice of 
atmospheric parameters.
Separate isotopic components of a line all share the same initial
and final energy levels (since the HFS of these
levels is negligible relative to the energy differences themselves), 
eliminating systematic errors associated with
the excitation and ionization states of each isotope.

\subsection{The Europium Isotopic Fraction}
\label{euiso}

\subsubsection{In the Sun}
\label{eusun}

For S.~S.\ material, the \ncap\ isotopic abundances 
are precisely known from studies of terrestrial meteorites.
The Eu~\textsc{ii} isotopic fraction in the Solar
spectrum was first studied by \citet{hauge70} and 
\citet{hauge72}, who found 
$^{153}$Eu/$^{151}$Eu\,=\,(48\,$\pm$\,6)/(52\,$\mp$\,6)
(or $f_{151}\,=\,0.52$ in our notation).
\citet{lawler01b} reevaluated the Solar Eu isotopic fraction,
finding $f_{151}\,=\,0.50\,\pm\,0.07$ from the 
4129\AA\ line.
We do not repeat this measurement here.

\subsubsection{In Metal-Poor Stars}
\label{eustars}

\begin{figure}
\epsscale{1.0}
\plotone{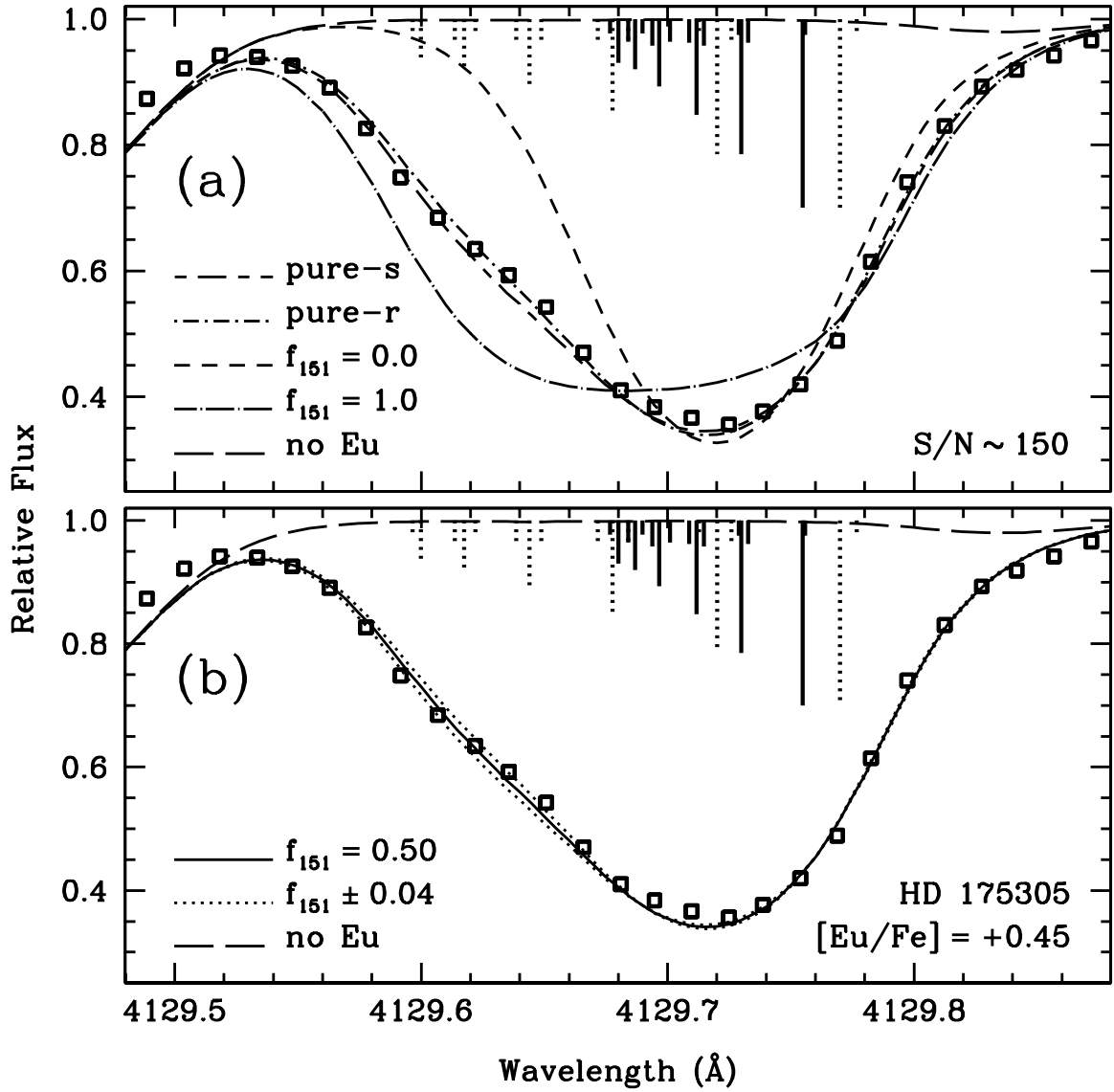}
\caption{
\label{fig4129iso} 
Our synthesis of the Eu~\textsc{ii} line at 4129\AA\
in HD~175305.
We have measured an isotopic fraction of
$f_{151}\,=\,0.50\,\pm\,0.04$ for this line.
The observed spectrum is indicated by open squares.
In panel~\textbf{a}, we show the 
syntheses with the extreme values of $f_{151}$
as well as the pure-$s$- and pure-\rpro\ syntheses.
In panel~\textbf{b}, 
the solid curve represents our best fit isotopic fraction, 
while the dotted curves represent the 3$\sigma$ uncertainties.
In both panels, 
the long-dashed curve is a synthesis with no Eu 
present, indicating that this line is generally
free of blending features. 
The vertical sticks represent the positions and 
relative strengths of the individual hyperfine
components; 
dotted sticks represent $^{151}$Eu while
solid sticks represent $^{153}$Eu.
Note that the Eu elemental abundance reported
is the best-fit value for this synthesis
and not our derived Eu abundance for HD~175305.
}
\end{figure}

\begin{figure}
\epsscale{1.0}
\plotone{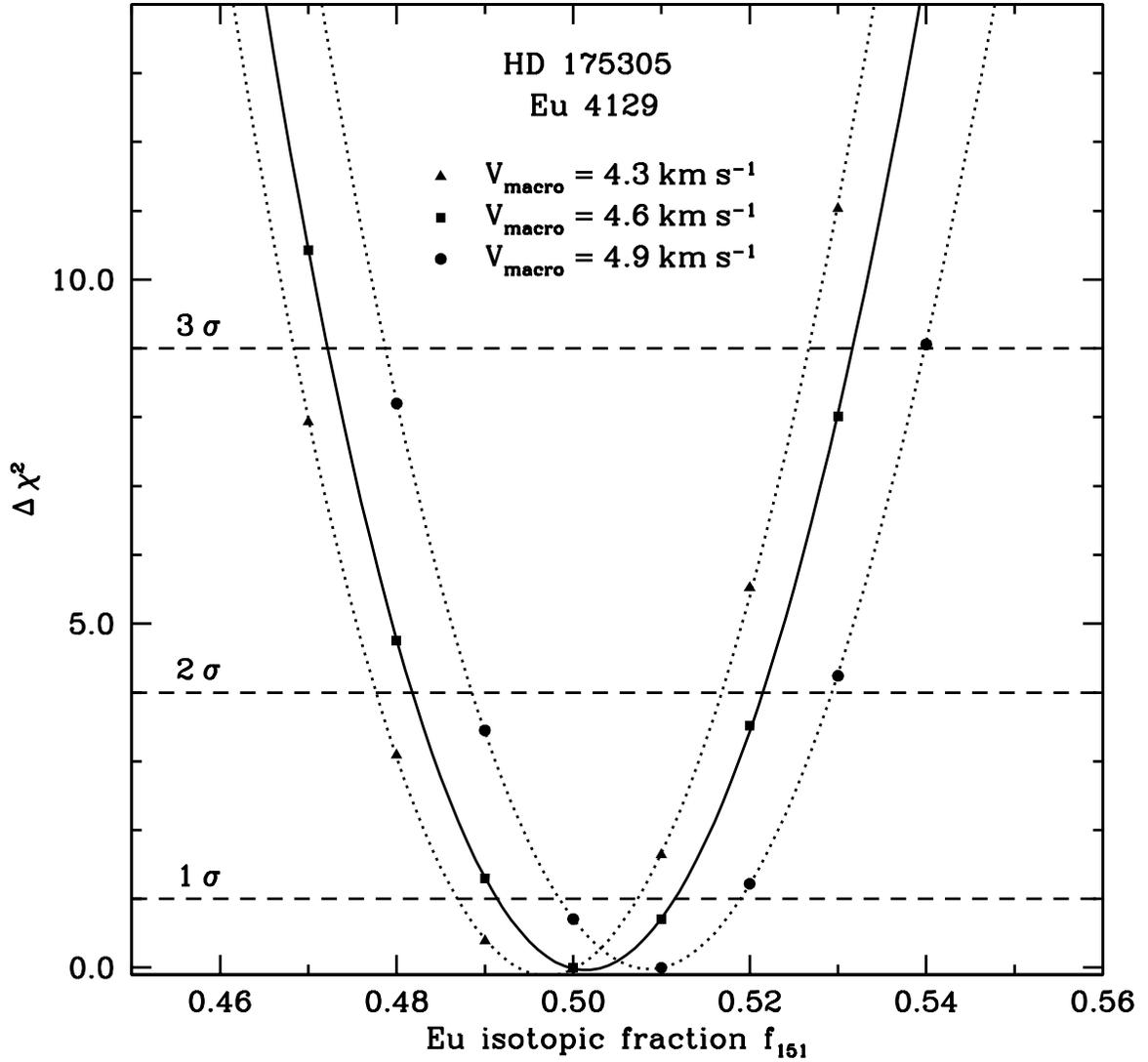}
\caption{
\label{4129chisq}
$\Delta\,\chi^{2}$ values for the Eu 4129 line in HD~175305,
shown for three values of $V_{\rm macro}$.
The solid line is a third-order polynomial fit to the measured
$\Delta\,\chi^{2}$ values for the best value of $V_{\rm macro}$,
4.6\,km\,s$^{-1}$, and the dotted lines are fits to the 
measured $\Delta\,\chi^{2}$ values for the uncertainties in 
$V_{\rm macro}$.
One, two, and three-$\sigma$ confidence intervals are indicated
by the dashed lines.
The fits have well-defined minima, and the choice of the
macroturbulent velocity has little effect on the derived isotopic fraction.
}
\end{figure}

\begin{figure}
\epsscale{1.0}
\plotone{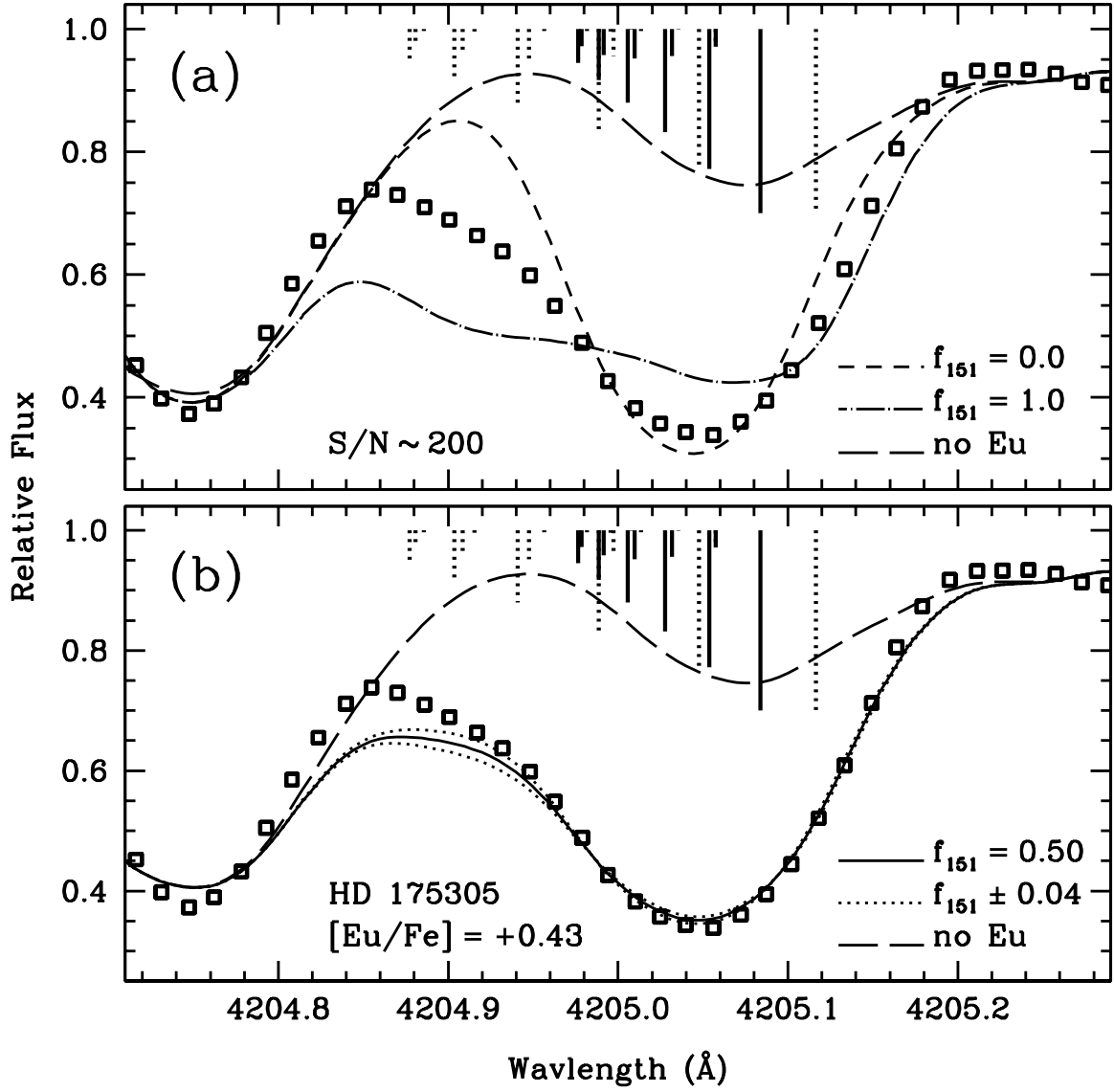}
\caption{
\label{fig4205iso} 
Our synthesis of the Eu~\textsc{ii} line at 4205\AA\
in HD~175305.
Symbols are the same as in Figure~\ref{fig4129iso}.
We emphasize that the isotopic fraction shown in 
panel~\textbf{b} is
the value derived from the 4129\AA\ line 
and not a fit to this line.
Note the strong blends.
The isotope fraction derived from the
4129\AA\ line provides a reasonable fit
to the 4205\AA\ line as well,
requiring only small changes in the elemental abundance.
}
\end{figure}

\begin{figure}
\epsscale{1.0}
\plotone{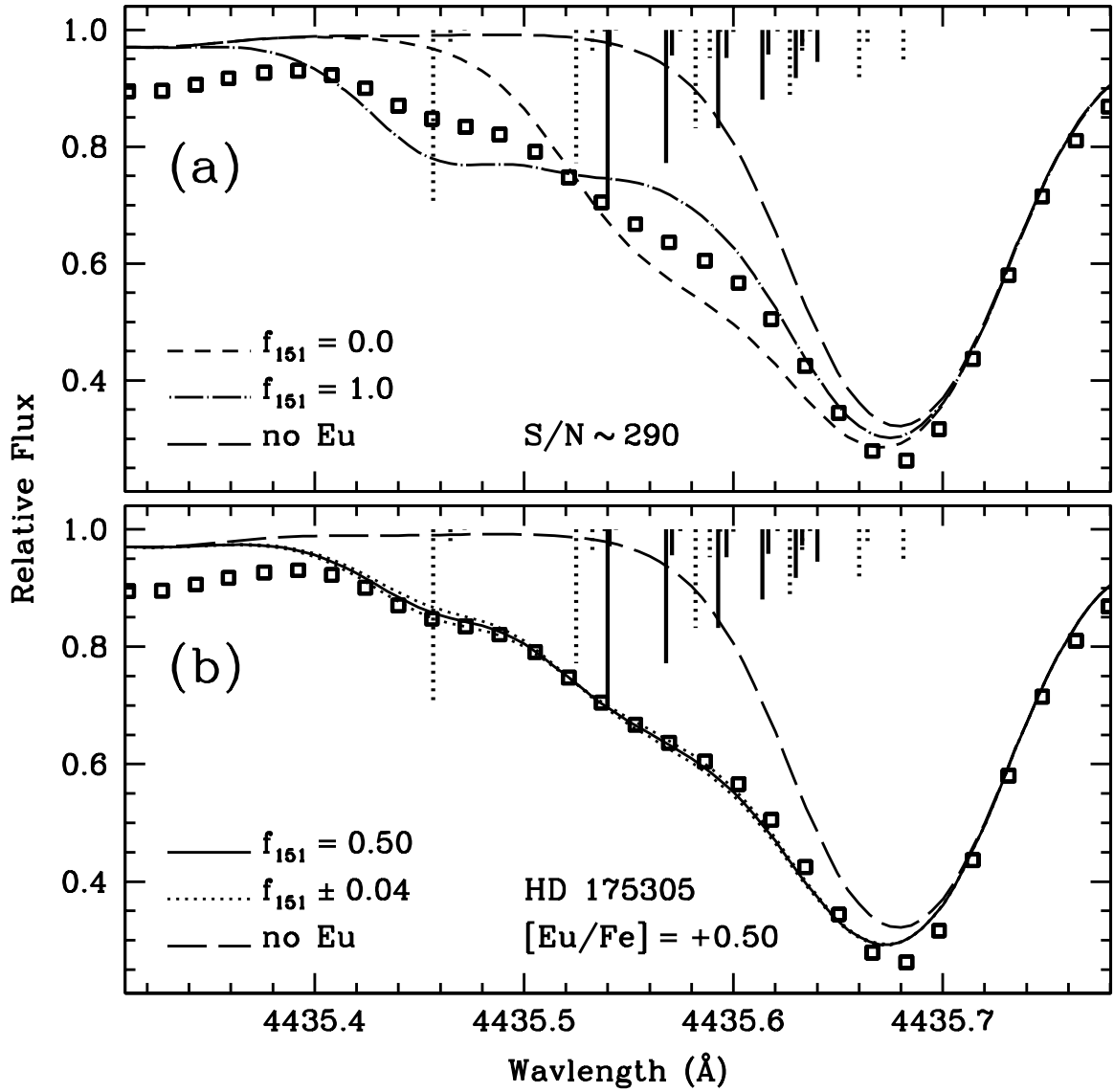}
\caption{
\label{fig4435iso} 
Our synthesis of the Eu~\textsc{ii} line at 4435\AA\
in HD~175305.
Symbols are the same as in Figure~\ref{fig4129iso}.
We emphasize that the isotopic fraction shown in 
panel~\textbf{b} is
the value derived from the 4129\AA\ line 
and not a fit to this line.
Note the strong blend with the saturated
Ca~\textsc{i} line at 4535.67\AA.
The isotope fraction derived from the
4129\AA\ line provides a reasonable fit
to the 4435\AA\ line as well,
requiring only small changes in the elemental abundance.
}
\end{figure}

From the Eu~\textsc{ii} 4129\AA\ line in HD~175305 
we measure an isotopic fraction
of $f_{151}\,=\,0.50\,\pm\,0.04$,
as shown in Figure~\ref{fig4129iso}.
The $\Delta\,\chi^{2}$ values are shown in Figure~\ref{4129chisq}
for different values of the isotopic fraction and 
macroturbulent velocity.
The minimum $\Delta\,\chi^{2}$ values have well-defined minima,
leading to the small uncertainties in the derived isotopic fraction.
As noted in Appendix~\ref{eulines}, we cannot trust 
our syntheses of the Eu~\textsc{ii} lines at 4205
and 4435\AA\
at the precision necessary to measure an isotopic fraction.
For the sake of comparison,
in Figures~\ref{fig4205iso} and \ref{fig4435iso}
we show our synthesis of these two lines in HD~175305
given our best estimate of the blending features,
but we emphasize that these syntheses 
display the best-fit isotopic fraction derived from
the 4129\AA\ line.

The Eu~\textsc{ii} 4129\AA\ line was not covered in our spectra of
either HD~196944 or HD~122563.

Although only two naturally-occurring isotopes of Eu exist,
only a small spread is predicted between 
$f^{s}_{151}$ and $f^{r}_{151}$ (see panel~\textbf{a} of 
Figure~\ref{fig4129iso}).
It is difficult even with very high-resolution ($R\,\gtrsim\,100,000$) 
and high-S/N ($\sim$~few hundred) data on one or more lines
to achieve sufficiently small uncertainties on isotopic abundances
to determine the nucleosynthetic source of the Eu.
\citet{sneden02,aoki03a,aoki03b} have successfully shown that
the Eu $s$- and \rpro\ nucleosynthetic signatures
can be distinguished at the isotopic level,
using approximately three lines in each of seven stars.
The measurements made from individual lines in each 
star in these studies have a large degree of scatter,
causing the final Eu isotopic fraction to be
mildly sensitive to the set of lines chosen.
For example, we note that  
\citet{aoki03a}'s two measurements of $f_{151}$ from
the 4435\AA\ line appear systematically higher
than the results obtained from the 4205\AA\ line
by $\approx$~10$-$15\%,
which is roughly the same as the change in the
Eu isotopic fractions
that are trying to be distinguished.
Although we do not measure an isotopic fraction
from the 4435\AA\ line in HD~175305, our syntheses
hint at a similar result in the small
region without obvious blends (4435.40$-$4435.50\AA).
Additional measurements of the Eu isotopic fraction
from these lines would be helpful to confirm or refute these 
possible systematic trends.
If such systematic trends exist, they likely
result from the presence of blending features,
and the measured isotopic fractions
may be as sensitive to the blending features
as to the Eu isotopes themselves.
There is little hope of being able to 
distinguish Eu isotopic fractions with any greater
precision than has already been achieved 
using these methods on this set of lines, including the
Eu~\textsc{ii} lines at 3819 and 3907\AA.

\subsection{The Samarium Isotopic Fraction}
\label{smiso}

\subsubsection{In the Sun}
\label{smsun}

\citet{ekeland75} analyzed the Sm isotopic fraction in the 
Solar photosphere from two lines, 4467 and 4519\AA, 
finding isotopic fractions consistent with the
meteoritic values.
The 4467\AA\ line is broadened slightly by HFS, with
almost zero broadening resulting from the isotope shifts,
while the
4519\AA\ line is broadened slightly by isotope shifts.
High-precision single-frequency laser measurements of the
isotope shifts for $^{147}$Sm and $^{149}$Sm
have not been measured in the lab for the 4519\AA\ line,
therefore we do not attempt synthesis of this line.

We synthesize the 4467\AA\ line using the \citet{holweger74}
model atmosphere and the \citet{kurucz84} Solar spectrum.
We fix the broadening from nearby Fe lines.
The total spread in the HFS and isotope shifts for this 
line are $\lesssim$~0.05\AA\ and $\lesssim$~0.01\AA, respectively, 
and we cannot measure the isotopic
fraction, either in terms of $f_{\rm odd}$ or $f_{152+154}$,
from the line-profile-fitting method
even when a high-quality Solar spectrum is used.
\citet{ekeland75} provide very little information about
their method of measuring the Sm isotopic fraction,
and their very small uncertainties ($f_{\rm odd}\,=\,0.28\,\pm\,0.07$)
for a line with relatively small HFS and isotope shifts
give us some reservation in accepting their result.
All of the other Sm transitions listed in 
Appendix~\ref{smlines} are either far too weak
or are severely-blended with strong transitions
of other species to measure reliable isotopic fractions
in the Solar spectrum, so we are unable to make
any measurements of the Sm isotopic fraction 
in the Solar photosphere.
We remind the reader that the blended lines may be reliable
for measurements of the Sm elemental abundance,
but we cannot trust our syntheses of the blending
features at a precision necessary to measure
isotopic fractions.

\subsubsection{In Metal-Poor Stars}
\label{smstars}

\begin{figure}
\epsscale{1.0}
\plotone{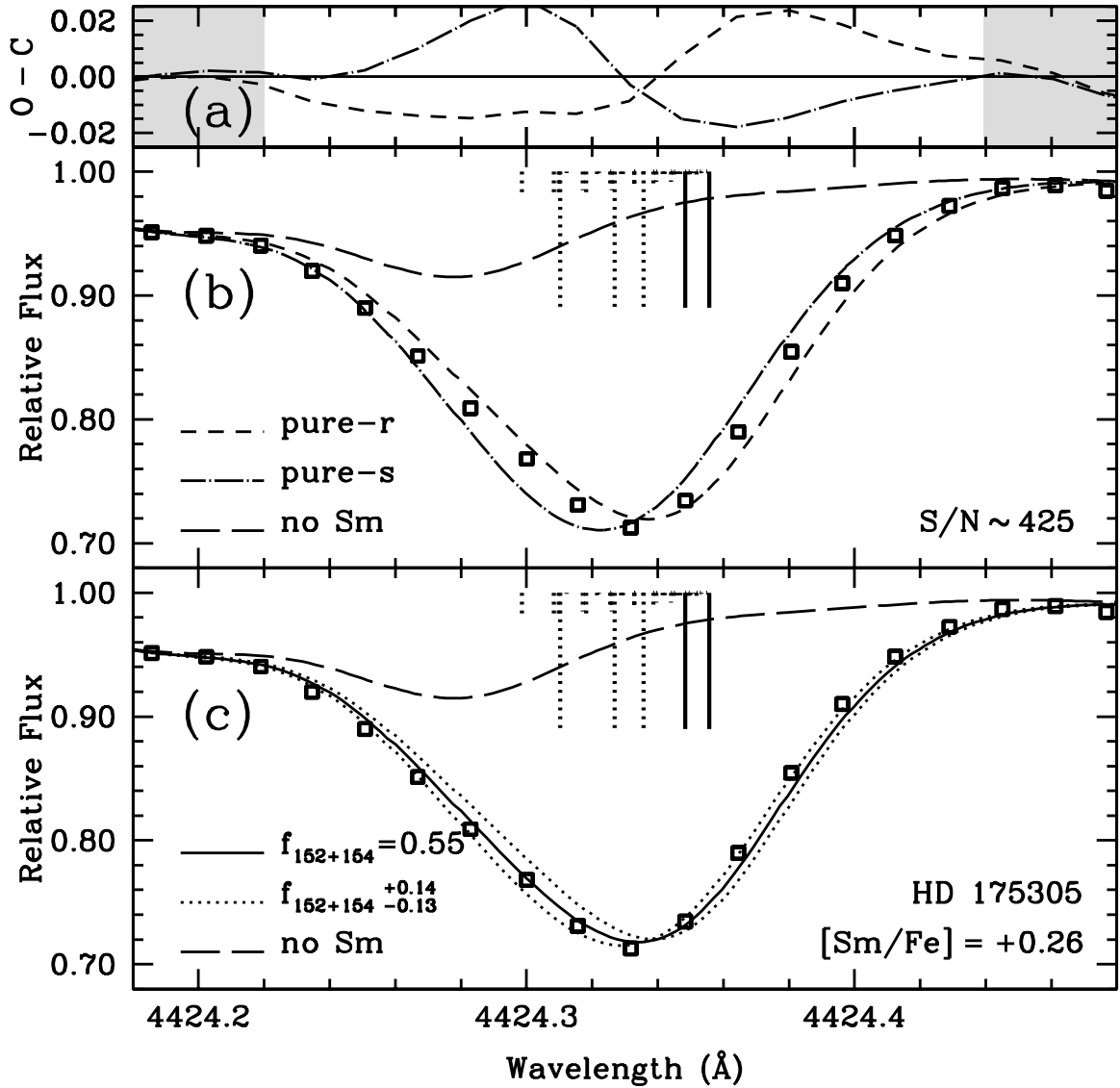}
\caption{
\label{4424iso175}
Our synthesis of the Sm~\textsc{ii} line at 4424\AA\
in HD~175305.
The observed spectrum is indicated by open squares.
In panel~\textbf{a}, we show an ($O-C$) plot
for the pure-$s$- and pure-$r$-process syntheses
shown in panel~\textbf{b}.
The unshaded region indicates the wavelengths
over which we measure the isotopic fraction
using the $\chi^{2}$ algorithm.
In panel~\textbf{b}, we show syntheses for the 
pure-$s$- and pure-$r$-process syntheses 
(the dot-dashed and short-dashed curves, respectively)
illustrating the contrast between these extremes.
In panel~\textbf{c}, we show the synthesis of our
best fit, represented by the solid curve, and the
3$\sigma$ uncertainties, represented by the dotted curves.
The long-dashed curve is a synthesis with no Sm
present, indicating the presence of blending features.
The vertical sticks represent the positions and 
relative strengths of the individual hyperfine components;
solid sticks represent the two heaviest isotopes,
while dashed sticks represent the lighter five isotopes.
We measure a Sm isotopic fraction of
$f_{152+154}\,=\,0.55^{+0.14}_{-0.13}$ for this line,
which suggests an $r$-process origin.
Note that the Sm elemental abundance reported is
the best-fit value for this line only and not 
our mean [Sm/Fe] in HD~175305.
}
\end{figure}

\begin{figure}
\epsscale{1.0}
\plotone{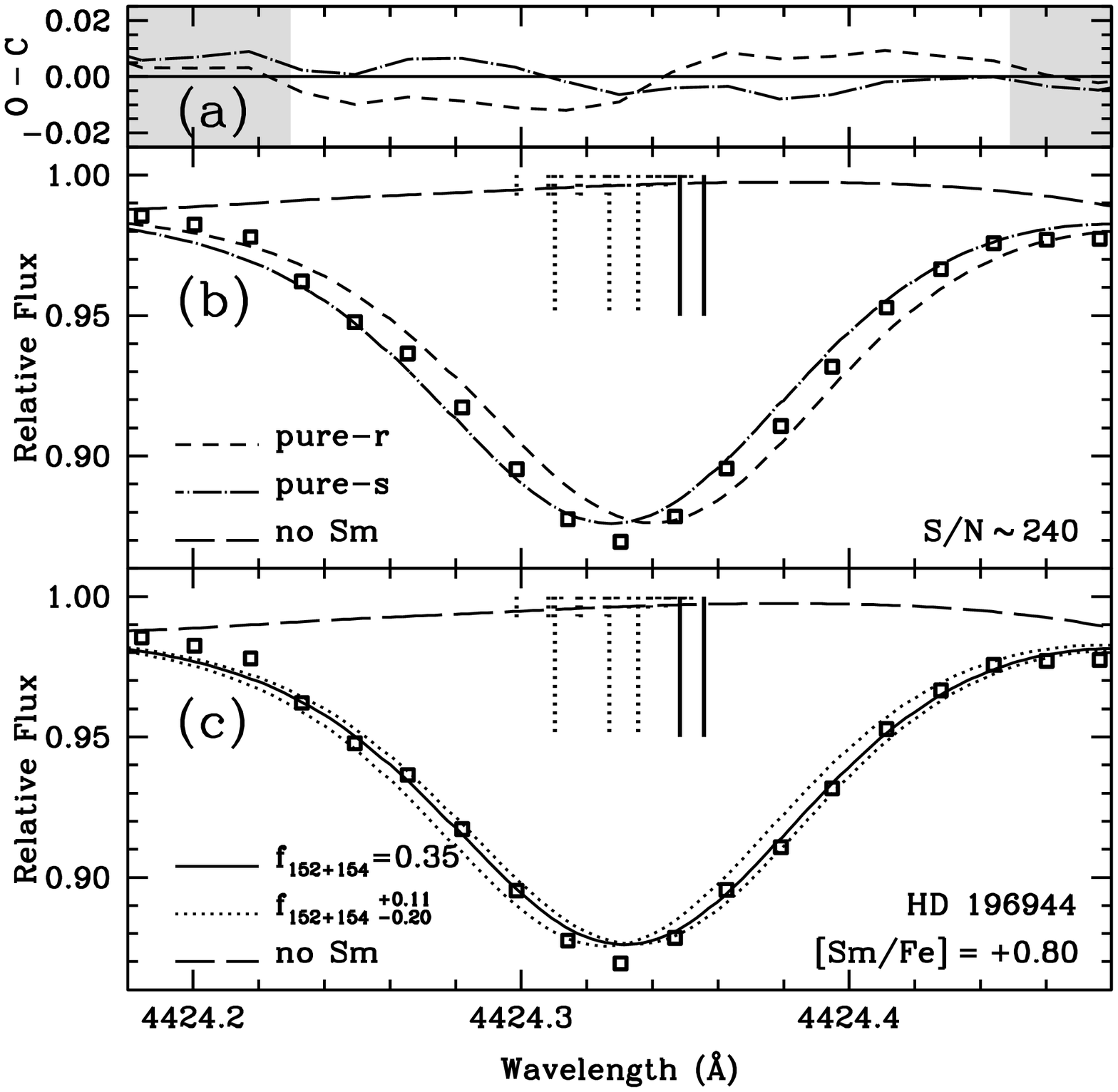}
\caption{
\label{4424iso196}
Our synthesis of the Sm~\textsc{ii} line at 4424\AA\
in HD~196944.
Symbols are the same as in Figure~\ref{4424iso175}.
We measure a Sm isotopic fraction of
$f_{152+154}\,=\,0.35^{+0.11}_{-0.20}$ for this line,
which moderately suggests an \spro\ origin.
Note that the Sm elemental abundance reported is
the best-fit value for this line only and not 
our mean [Sm/Fe] in HD~175305.
}
\end{figure}

Figures~\ref{4424iso175} and \ref{4424iso196} 
show our syntheses of the Sm~\textsc{ii}
line at 4424\AA\ in HD~175305 and HD~196944, respectively,
and the results from our line profile fits.
We also include the Cr abundance of the 
Cr~\textsc{i} 4424.28\AA\ line, which blends
the blue wing of the Sm line, as a free parameter. 
In HD~122563, this Sm line has a relative depth of 
$\sim$\,2\%.
We do not claim measurements of either the elemental abundance
or isotopic fraction of Sm from this line in HD~122563;
it is clear that this line is simply too weak to 
warrant further analysis.
In HD~175305, we measure a Sm isotopic fraction of
$f_{152+154}\,=\,0.55\,\pm\,0.14$.
In HD~196944, we measure a Sm isotopic fraction of
$f_{152+154}\,=\,0.35\,^{+0.11}_{-0.20}$.
In addition to showing the best fit curves and their
uncertainties in Figures~\ref{4424iso175} and \ref{4424iso196},
we also show the syntheses for pure-$s$- and pure-$r$-process
nucleosynthesis and the ($O-C$) curves for these syntheses, 
illustrating the contrast between the two extremes.
In each of these figures, the pure-$s$- and pure-$r$-process
syntheses in panel~\textbf{b} are compared to the
observed spectra in panel~\textbf{a},
illustrating the degree to which the observed spectrum
tends toward one synthesis or the other.

\begin{figure}
\epsscale{1.0}
\plotone{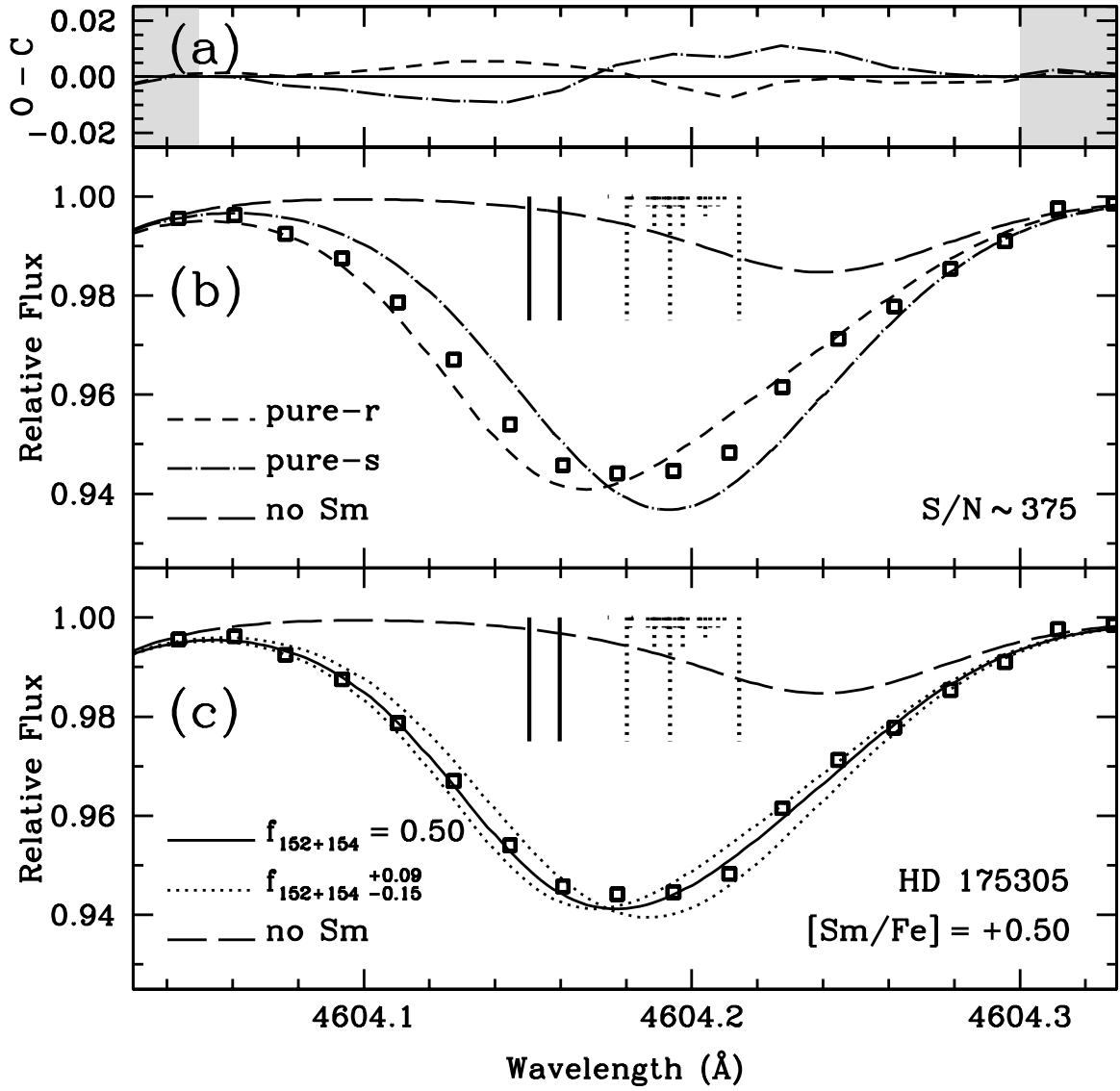}
\caption{
\label{4604iso}
Our synthesis of the Sm~\textsc{ii} line at 4604\AA\ 
in HD~175305.  
Symbols are the same as in Figure~\ref{4424iso175}.
We measure a Sm isotopic fraction of
$f_{152+154}\,=\,0.50^{+0.09}_{-0.15}$ from this line,
which suggests an \rpro\ origin.
Note that the Sm elemental abundance reported is
the best-fit value for this line only and not 
our mean [Sm/Fe] in HD~175305.
}
\end{figure}

\begin{figure}
\epsscale{1.0}
\plotone{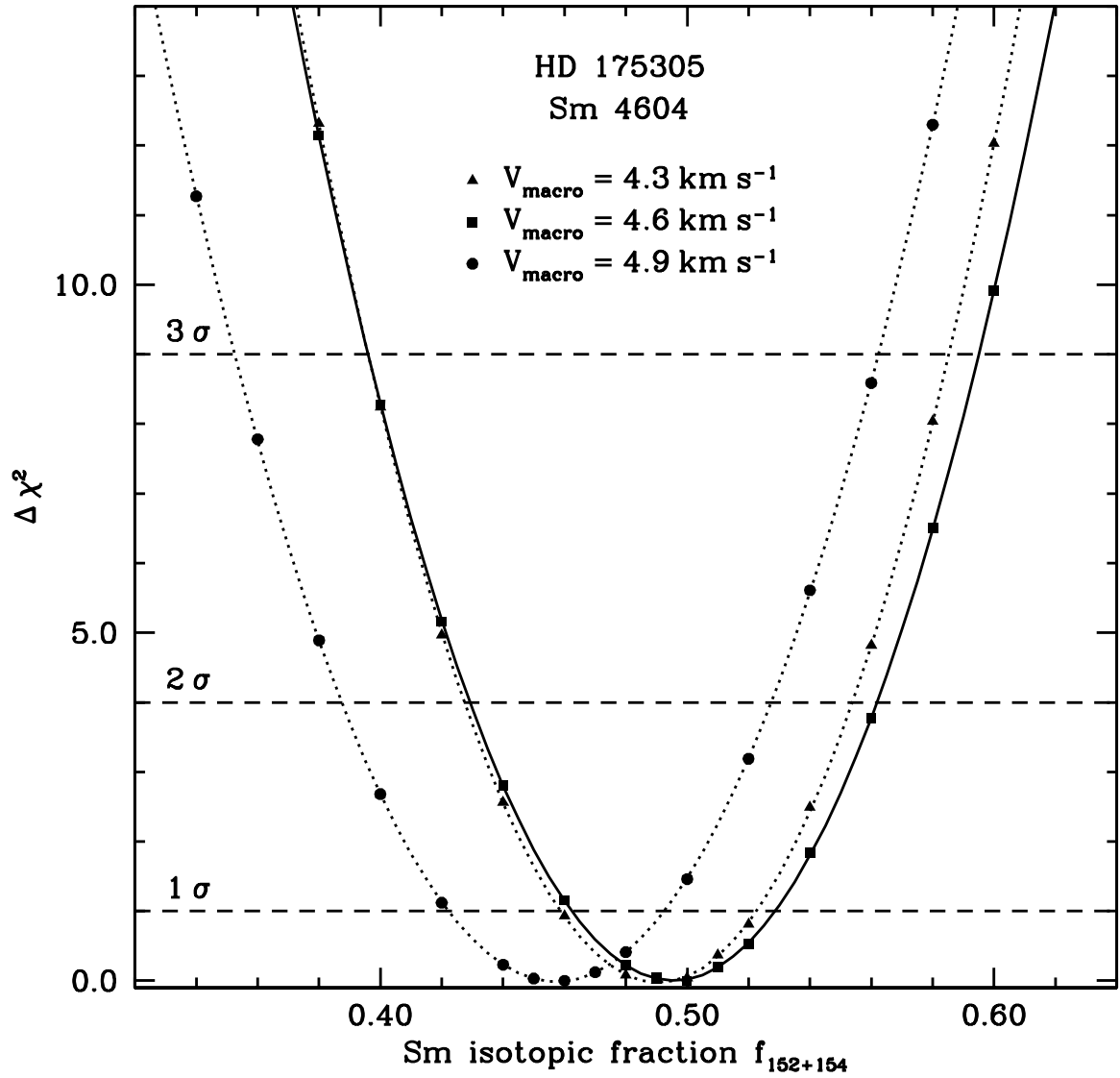}
\caption{
\label{4604chisq}
$\Delta\,\chi^{2}$ values for the Sm 4604 line in HD~175305,
shown for three values of $V_{\rm macro}$.
Symbols are the same as in Figure~\ref{4129chisq}.
Here the macroturbulence has only a small effect on the 
derived isotopic fraction.
Similar results are found for the Sm 4424 line in both
HD~175305 and HD~196944.
}
\end{figure}

Figure~\ref{4604iso} shows our synthesis of the
Sm~\textsc{ii} line at 4604\AA\ in HD~175305.
We measure a Sm isotopic fraction of
$f_{152+154}\,=\,0.50^{+0.09}_{-0.14}$.
This line is weak, with a relative depth
of only 5\%.
Figure~\ref{4604chisq} shows the $\Delta\,\chi^{2}$ values
for our fit to this line; $\Delta\,\chi^{2}$ values 
from the Sm 4424\AA\ line
in both stars are similar to the 4604\AA\ line.

If instead we match the observed and synthetic spectra at the point
of insensitivity to the isotopic fraction, we find
$f_{152+154}\,=\,0.41\,\pm\,0.32$ from the 4424\AA\ line,
$f_{152+154}\,=\,0.49\,\pm\,0.17$ from the 4604\AA\ line, and
$f_{152+154}\,=\,0.49\,\pm\,0.20$ from the 4719\AA\ line in HD~175305.
In HD~196944, we find
$f_{152+154}\,=\,0.37\,\pm\,0.28$ from the 4424\AA\ line.
While we regard the line profile fits as the superior measurements
of the isotopic fraction, the agreement between the sets of measurements
from the two methods is reassuring.

\subsection{The Neodymium Isotopic Fraction}
\label{ndiso}

\subsubsection{In the Sun}
\label{ndsun}

We check our Nd isotopic fraction measurements 
against the Solar spectrum.
We are aware of no previous attempts to examine the
Nd isotopic mix in the Solar photosphere, but the
individual S.~S.\ isotopic abundances have been 
established from studies of terrestrial meteorites.
The Nd~\textsc{ii} line at 4446\AA\ would appear to be
the most reliable indicator of the Nd isotopic fraction
due to its relative strength and freedom from blending features.
The line profile shape does not change enough to allow 
us to apply our fitting algorithm to measure the isotopic fraction, and
we can find no good candidate Fe~\textsc{i} or Ti~\textsc{ii} 
lines nearby to match the observed and synthetic spectra.
By using the method of matching the observed and synthetic 
spectra at the point of insensitivity to the isotopic mix,
we derive an isotopic fraction of
$f_{142+144}\,=\,0.53^{+0.47}_{-0.53}$, 
which is unfortunately unable to exclude any possible 
combination of isotopes.
Therefore, we are unable to provide independent 
confirmation of the S.~S.\ Nd isotopic fraction
from analysis of the Solar photosphere.

\subsubsection{In Metal-Poor Stars}
\label{ndstars}

Because the isotope shifts of Nd are smaller than Sm and 
because we cannot match our observed and synthetic spectra
with a high absolute accuracy,
we can only rely on the third matching method to measure
the isotopic fraction of the Nd lines in these stars.
In HD~175305, we measure 
$f_{142+144}\,=\,0.25^{+0.75}_{-0.25}$ from the 4177\AA\ line,
$f_{142+144}\,=\,0.11^{+0.85}_{-0.11}$ from the 4446\AA\ line, and
$f_{142+144}\,=\,0.33^{+0.67}_{-0.33}$ from the 4567\AA\ line.
The mean isotopic fraction of these three lines is
$f_{142+144}\,=\,0.21^{+0.56}_{-0.21}$.
Combining these measurements together using the asymmetric
uncertainties that reflect the fact that the isotopic fraction
cannot be less than 0.0 or greater than 1.0
could introduce a bias into the weighted mean.
Therefore we have estimated the uncertainties in these measurements
also by assuming (non-physically) that the isotopic fraction
can exceed these limits.
We find uncertainties of $\pm\,1.10$, $\pm\,0.85$, and $\pm\,1.00$
on these three lines, respectively, resulting in a
weighed uncertainty of $\pm\,0.56$ on the mean.
Our final quoted uncertainties do impose the physical limits 
on the isotopic fractions.
In HD~196944, we measure
$f_{142+144}\,=\,0.36^{+0.64}_{-0.36}$ from the 4446\AA\ line.
The non-physical uncertainty on this measurement is $\pm\,0.68$,
which provides a relative comparison to our ability to measure
the 4446\AA\ line in HD~175305.
Unfortunately none of these individual measurements 
provides any real insight into the Nd isotopic fraction 
due to the size of the uncertainties.
The large uncertainties also prevent us from pointing out 
even a contrast between the Nd isotopic mixes in the two stars,
much less make any detailed assessment of the \spro\ 
nucleosynthetic predictions.

\section{Interpretation of the Nucleosynthetic Signatures}
\label{interpret}

In HD~175305,
our Sm isotopic fraction
$f_{152+154}\,=\,0.51\,\pm\,0.08$, a weighted average of our
measurements from profile fits to the 4424 and 4604\AA\ lines,
suggests an \rpro\ origin.
It is tempting to surmise that our Nd isotopic fraction, 
$f_{142+144}\,=\,0.21^{+0.56}_{-0.21}$,
derived from three lines, is suggestive of an 
\rpro\ origin, but the large uncertainty
cannot exclude an \spro\ origin.
Our Eu isotopic fraction,
$f_{151}\,=\,0.50\,\pm\,0.04$,
derived from only the 4129\AA\ line
and shown to be a reasonable match for the
4205 and 4435\AA\ lines,
implies an \rpro\ origin by the \citet{cowan06a} predictions but 
excludes neither a pure-$s$- nor a pure-$r$-process origin
by the \citet{arlandini99} predictions.
It is interesting to note that our isotopic fractions in 
HD~175305 suggest a nucleosynthetic history 
predominantly---but perhaps not purely---dominated by 
the \rpro, which also is suggested by the elemental
abundance trends.
The fact that both the elemental and the isotopic distributions 
in HD~175305 agree with the predicted elemental and isotopic distributions 
of \rpro\ material in the S.~S.\ adds to the 
preponderance of evidence that supports the hypothesis of a universal 
\rpro\ mechanism for elements with $Z\geq56$.

In HD~196944, 
our measured Sm isotopic fraction,
$f_{152+154}\,=\,0.35\,\pm\,0.14$,
is suggestive of an \spro\ origin.
Our Nd isotopic fraction,
$f_{142+144}\,=\,0.36^{+0.64}_{-0.36}$
can exclude no possible values of the isotopic mix.
The interpretation for the Sm is consistent with
the \spro\ elemental abundance
signatures and radial velocity
variation described in \S~\ref{radvel}.

\begin{figure}
\includegraphics[angle=270,scale=0.65]{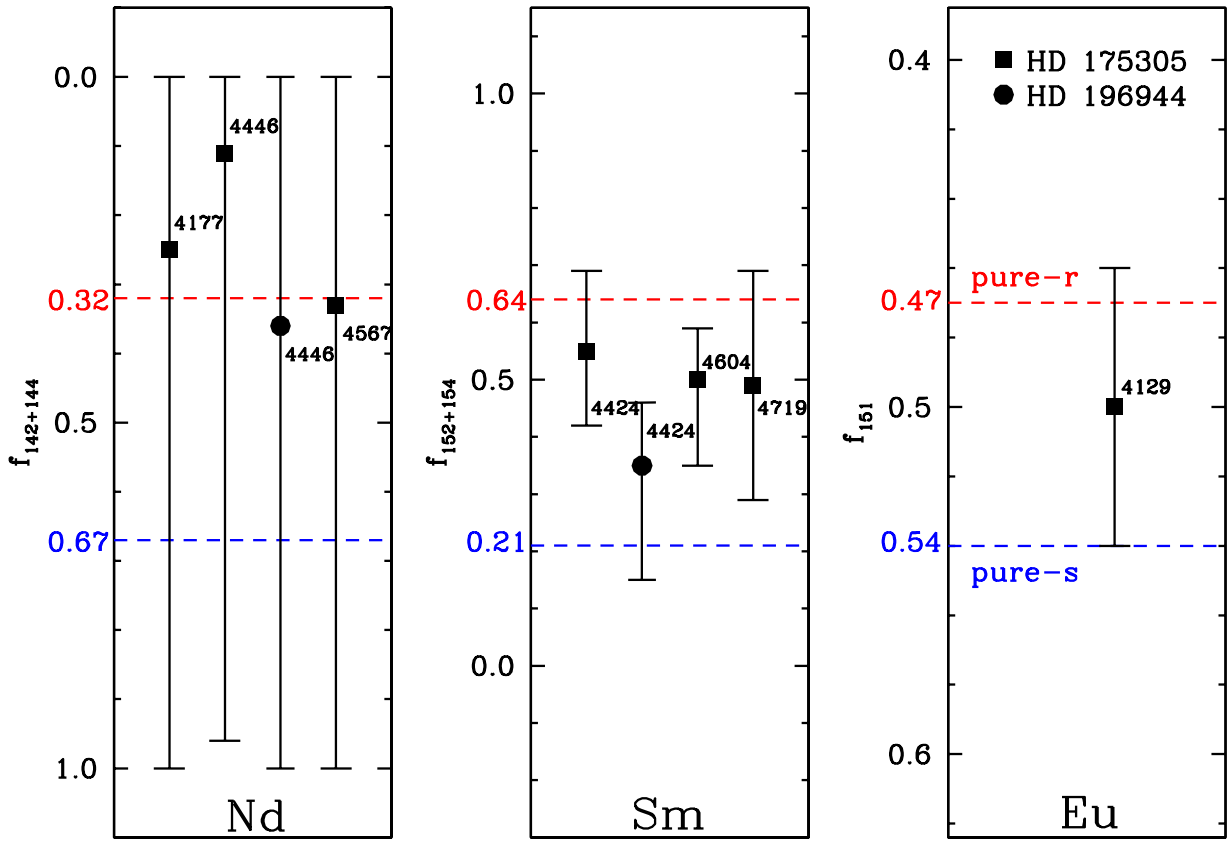}
%% B/W version should appear in print edition 
\caption{
\label{summary1}
A summary of our best isotopic fraction measurements for each species.
The Nd isotopic fractions reported here were measured 
by the method of matching the observed and synthetic spectra 
at the point of insensitivity to the isotopic mix.
The Sm and Eu isotopic fractions were measured by fitting the 
shape of the line profile.
Squares represent measurements in HD~175305, while circles represent
measurements in HD~196944.
The dotted line and dashed line (colored blue and red, respectively,
in the electronic edition) represent 
the pure-$s$- and pure-$r$-process predictions of \citet{arlandini99}.
The differences between the pure-$s$- and pure-$r$-process predictions
are scaled together to emphasize the relative precision with which 
the two processes can be distinguished by each species.
[See electronic edition for a color version of this figure.]
}
\end{figure}

\begin{figure}
\includegraphics[angle=270,scale=0.65]{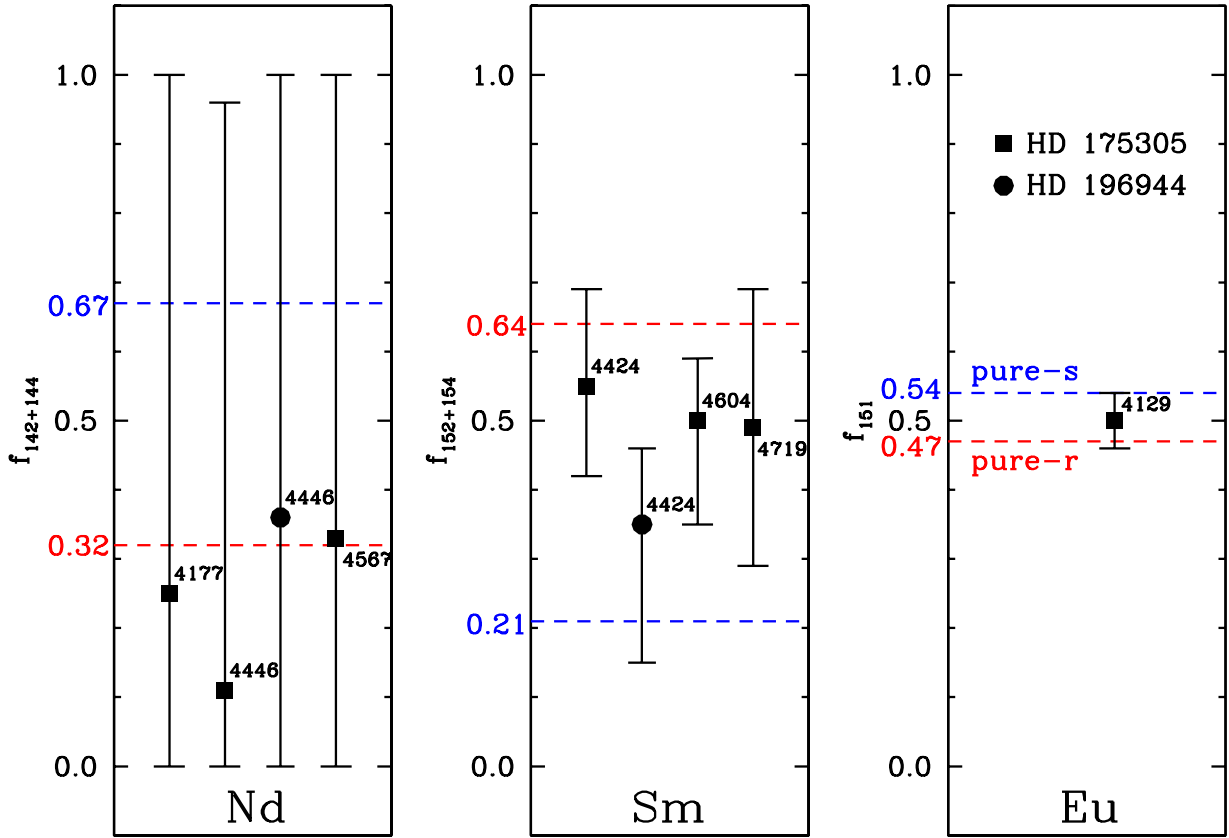}      
%% B/W version should appear in print edition 
\caption{
\label{summary2}
A summary of our best isotopic fraction measurements for each species.
The data and symbols are the same as in Figure~\ref{summary1}.
The measurements are cast here in an absolute sense, emphasizing
the relative precision with which each species was measured.
[See electronic edition for a color version of this figure.]
}
\end{figure}

Our best measurements for each species 
are summarized in Figures~\ref{summary1} and \ref{summary2}
for both HD~175305 and HD~196944.
In Figure~\ref{summary1}, the differences between the 
pure-$r$- and pure-$s$-process predictions are scaled together
to emphasize the relative precision with which the two processes can 
be distinguished by each species.
In contrast, Figure~\ref{summary2} displays our measurements 
in an absolute sense to emphasize the relative precision with 
which each species was measured. 
Expressed this way, it is clear that Eu can be measured precisely
but does not easily reveal its nucleosynthetic origin
because of the small difference between $f^{r}_{151}$ and
$f^{s}_{151}$.
The uncertainties on each Nd line can hardly exclude any 
possible value of the isotopic fraction.
Only the Sm isotopic fractions, which have uncertainties on the 
individual measurements that are smaller than the difference
between $f^{r}_{152+154}$ and $f^{s}_{152+154}$, 
offer any opportunity to distinguish the two processes 
with reasonable precision.

\section{Eu/Pb and $^{151}$Eu/Pb Abundance Ratios}
\label{eupb}

In \S~\ref{euatomic} we pointed out the great discrepancy
between the stellar model of \citet{arlandini99} and 
the classical method of \citet{cowan06a} in predicting the
amount of $^{151}$Eu produced by the \spro:
$N_{s}\,=\,0.00304\,\pm\,0.00013$ and
$N_{s}\,=\,0.0001$, respectively.
The stellar model predicts that the \spro\ should produce 
$\sim$~30 times more $^{151}$Eu than the classical method does.
For a certain amount of $^{151}$Eu produced by an \spro\
nucleosynthesis event, 
the classical method would then predict that the isotopes
near the termination of the $s$-process---such as Pb---should be 
overproduced by a factor $\sim$~30 relative to the stellar model.

We choose Pb for this comparative analysis because it is 
so dominantly produced by the \spro.  
In Pb-enriched metal-poor stars, such as HD~196944,
the Pb overabundance often will
reach [Pb/Fe]~$\geq +2.0$, making it easy to detect
at the Pb~\textsc{i} 4057\AA\ transition.
Because Eu is produced more easily by the \rpro\ than by the \spro\
(relatively speaking)
and Pb is produced more easily by the \spro\ than by the \rpro, 
the Eu/Pb and $^{151}$Eu/Pb abundance ratios in metal-poor 
stars should be extremely good discriminators of the 
predicted abundances of these species.

\citet{arlandini99} do not predict pure-$r$-process
Pb abundances due to the fact that Pb can be overproduced relative
to the lighter-$s$-process nuclei in low metallicity AGB stars.
In other words, at lower metallicity a greater number of neutrons is 
captured per Fe-peak seed than at higher metallicity, so the
\spro\ abundance distribution at low metallicity 
favors the heavest \spro\ nuclei; see, e.g., 
\citet{gallino98}, \citet{busso99}, and \citet{straniero06}.

\begin{figure}
\epsscale{1.0}
\plotone{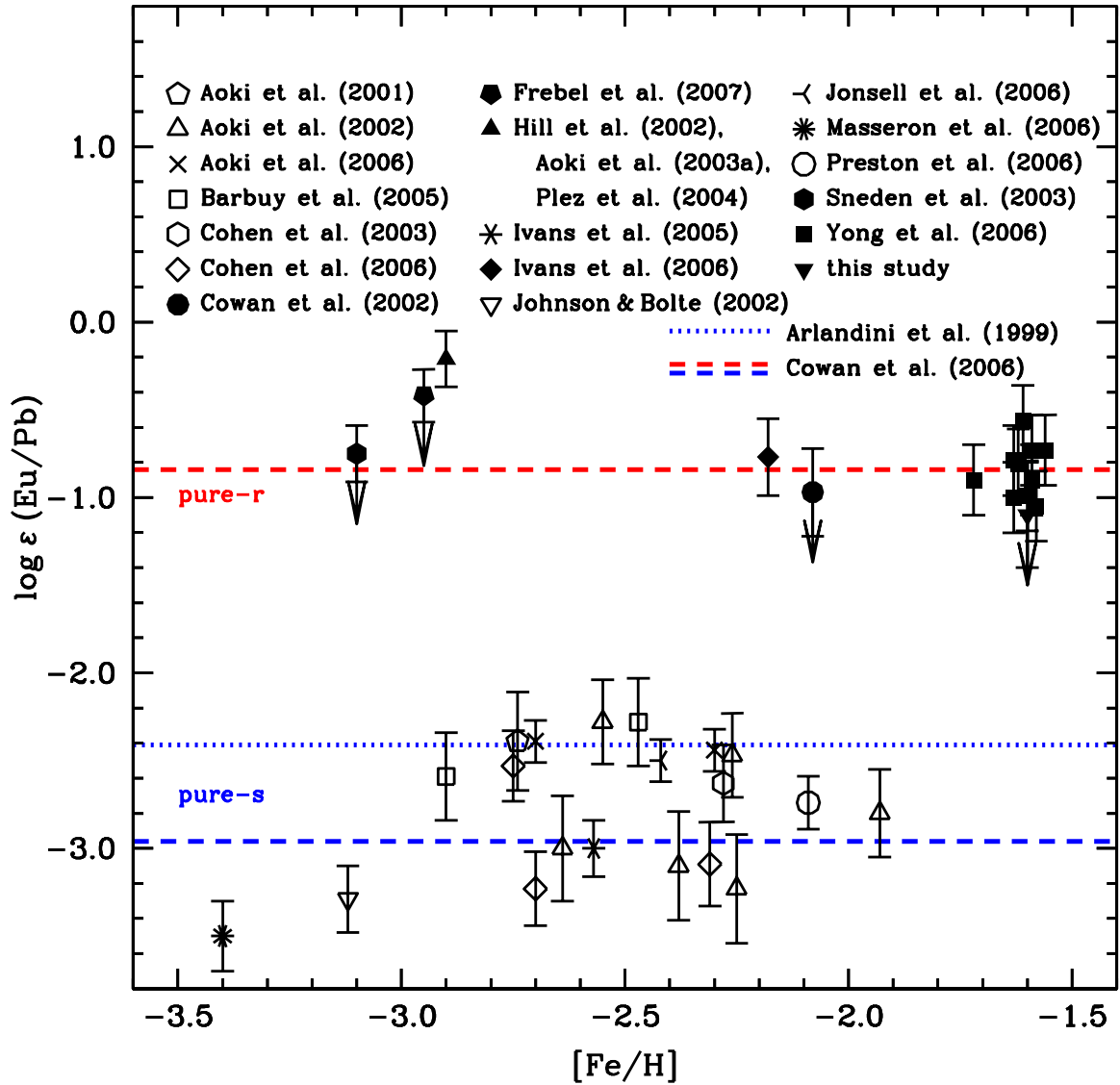}   %% B/W version should appear in print edition 
\caption{
\label{eupbplot}
The $\log\,\varepsilon\,$(Eu/Pb) abundance as a function of [Fe/H]
for 36 halo and globular cluster stars collected from the literature
or measured in our study.
References for the individual stars are indicated in the figure.
Solid shapes indicate stars classified as $r$-process enriched,
open shapes indicate stars classified as $s$-process enriched, and
starred shapes indicate stars classified as $(r+s)$-process enriched.
Upper limits are indicated with downward arrows.
The abundance predictions from \citet{arlandini99} and \citet{cowan06a} 
are shown as dotted and dashed lines, respectively.
(In the electronic edition, blue and red lines represent
the pure-$s$-process and pure-$r$-process abundance predictions.)
The $s$- and $(r+s)$-process-enriched stars cannot unambiguously distinguish
between the different pure-$s$ abundance predictions.
[See electronic edition for a color version of this figure.]
}
\end{figure}

In Figure~\ref{eupbplot} we have compiled the derived Eu/Pb 
ratios (or upper limits) for 35 metal-poor stars
from a variety of recent studies and for HD~175305.
The Eu/Pb ratio shows no obvious trends when expressed 
as a function of [Fe/H],
but we note that this sample of Eu and Pb abundances 
is far from complete.
Adopting the original authors' classifications for the
enrichment patterns for these stars, this sample consists
of 6 $r$-process-enriched stars 
(\citealt{cowan02,hill02,aoki03a,sneden03,plez04,ivans06,frebel07}; 
A.~Frebel, 2007, private communication), 
15 $s$-process-enriched stars
\citep{aoki01,aoki02,johnson02c,cohen03,barbuy05,cohen06,preston06},
and 5 $(r+s)$-process-enriched stars 
\citep{ivans05,aoki06,jonsell06,masseron06}
from the Galactic halo population.
The $r$-process-enhanced sample also includes 10 stars from 
the metal-poor globular clusters M\,13 and NGC\,6752
\citep{yong06}.
The small number of $r$-process enriched stars results from
the observational difficulty of detecting Pb in stars without
\spro\ enrichment.
The Eu/Pb ratios of the $r$-process-enhanced halo stars 
and globular cluster stars
all suggest that the \rpro\ predictions are correct.
Despite the fact that $\log\,\varepsilon\,$(Eu/Pb)$_{s}$ differs
by $\sim$~0.6~dex between the stellar model and classical method
predictions, the measured abundance spread in the 
$s$-process-enriched stars is still too great to unambiguously 
distinguish the two predictions from this information alone.
The $s$- and $(r+s)$-enhanced classes of stars show little difference from 
one another when their Eu and Pb abundances are expressed this way.

\begin{figure}
\epsscale{1.0}
\plotone{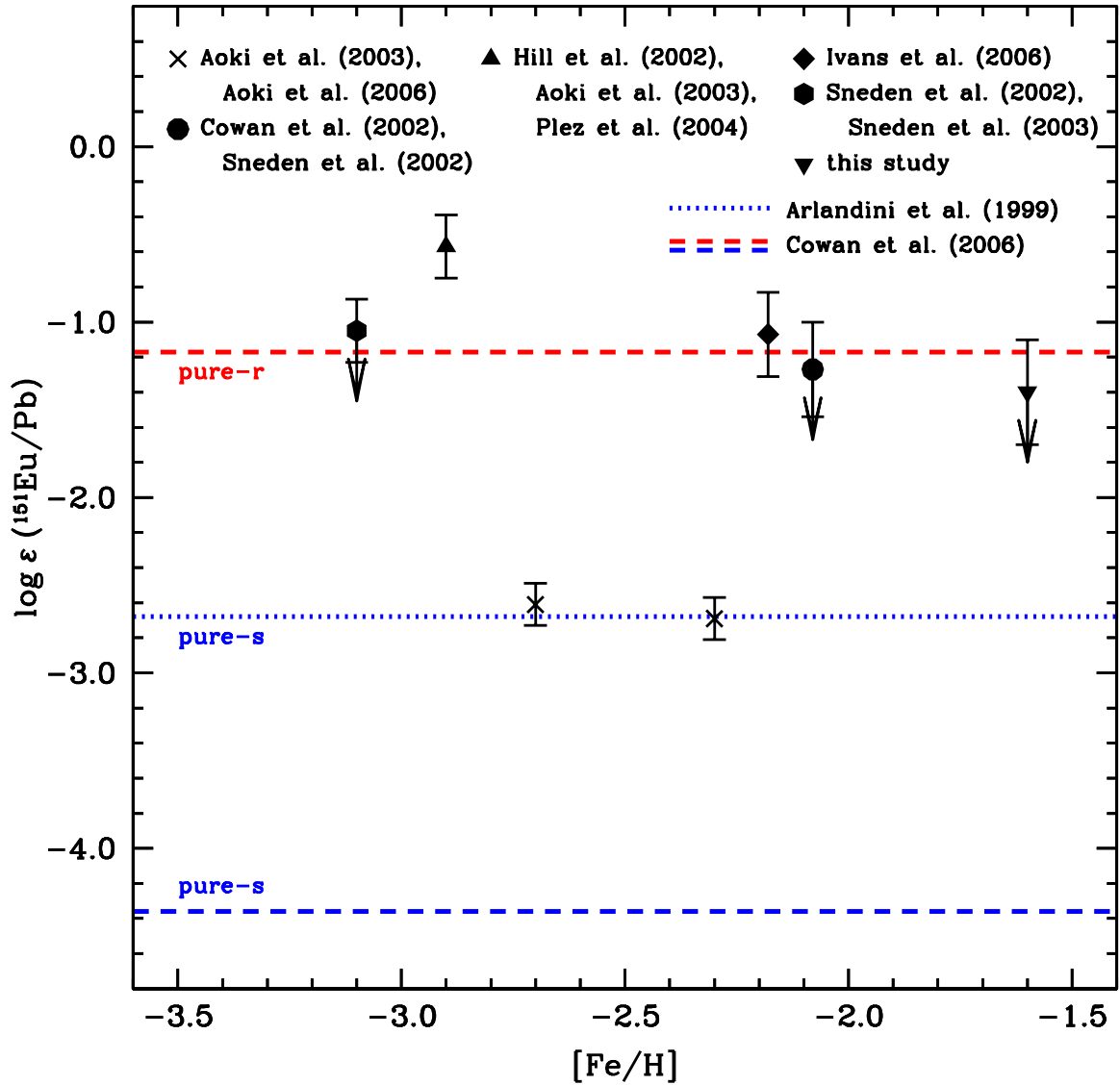}   %% B/W version should appear in print edition 
\caption{
\label{151pbplot}
The $\log\,\varepsilon\,$($^{151}$Eu/Pb) abundance as a function of [Fe/H]
for 7 halo stars collected from the literature or measured in our study.
References for the individual stars are indicated in the figure.
Symbols are the same as in Figure~\ref{eupbplot}.
The two $(r+s)$-enriched stars favor the pure-$s$ $^{151}$Eu 
abundance prediction of \citet{arlandini99}.
[See electronic edition for a color version of this figure.]
}
\end{figure}

In Figure~\ref{151pbplot} we have complied a subset of 
7 of these stars with measured Eu isotopic fractions, and 
we plot $\log\,\varepsilon\,$($^{151}$Eu/Pb) as a function of
metallicity.
This subset includes 5 $r$-process-enhanced stars and 
2 $(r+s)$-process-enhanced stars.
Again, the $r$-process-enhanced halo stars cluster around
the \rpro\ predictions.
Here, though, the stellar model and classical method predictions 
for $\log\,\varepsilon\,$($^{151}$Eu/Pb) differ by $\sim$~1.7~dex,
and the two $r+s$ stars with measured Eu isotopic fractions 
clearly favor the \citet{arlandini99} predicted $^{151}$Eu abundance.
This is not unexpected, given that \citet{aoki03a}, who measured
the Eu isotopic fraction in these two stars,
reported that $f_{151}$ was consistent with the 
\citet{arlandini99} predictions.

\begin{figure}
\epsscale{1.0}
\plotone{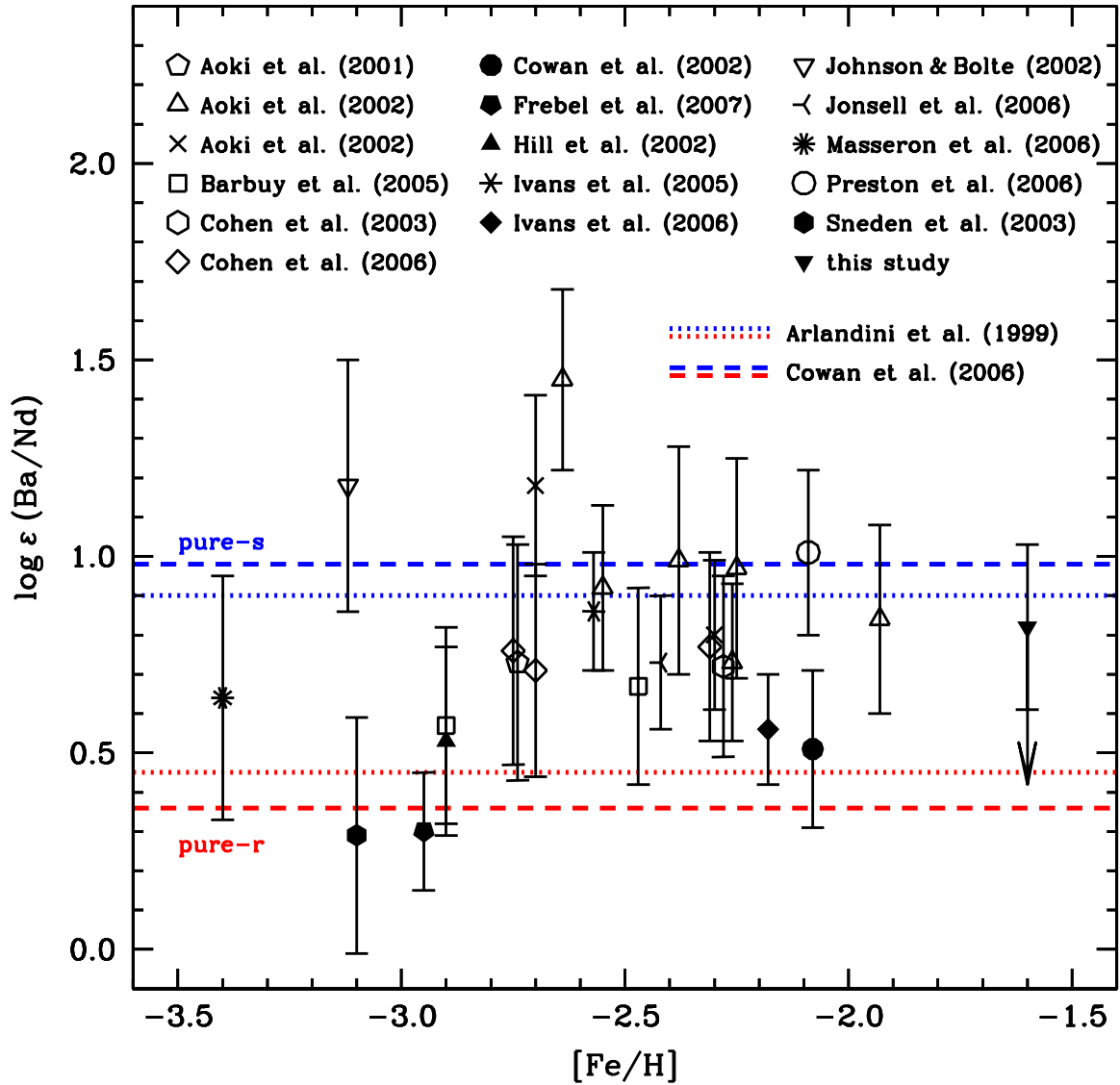}   %% B/W version should appear in print edition 
\caption{
\label{bandplot}
The $\log\,\varepsilon\,$(Ba/Nd) abundance as a function of [Fe/H]
for 26 halo stars collected from the literature or measured in our study.
References for the individual stars are indicated in the figure.
Symbols are the same as in Figure~\ref{eupbplot}.
No distinction is observed between the $s$- and $(r+s)$-enriched stars
from their Ba and Nd abundances, leading us to conclude that the 
two $(r+s)$-enriched stars in Figure~\ref{151pbplot} are representative
of stars containing products of predominantly-$s$-process-enriched material.
[See electronic edition for a color version of this figure.]
}
\end{figure}

It is worth considering whether these $r+s$ stars may have enough
\rpro\ enrichment to give them a $\log\,\varepsilon\,$($^{151}$Eu/Pb)
abundance that lies between the \citet{cowan06a} pure-$s$ and
pure-$r$ predictions, inadvertently suggesting the \citet{arlandini99}
value.
To investigate this matter, in Figure~\ref{bandplot} we plot the
$\log\,\varepsilon\,$(Ba/Nd) abundance ratios for the full sample 
of halo stars;
\citet{yong06} did not measure Ba or Nd abundances in their 
globular cluster stars.
While there is still a relatively large amount of scatter among
the $s$- and $(r+s)$-enriched stars, neither population appears
distinct from the other, and the two $r+s$ stars shown in 
Figure~\ref{151pbplot} scatter above and below the predicted 
pure-$s$-process abundance value.
We therefore conclude that these two stars are representative
of predominantly-$s$-process-enriched material, and these 
observations thereby support the \citet{arlandini99} 
$^{151}$Eu abundance prediction.
We caution that this conclusion hinges on the Eu isotopic
fraction measured in only two stars, and 
additional observations of the Eu isotopic fraction in 
$s$-process-enriched stars would be most helpful to further 
resolve the matter.

\citet{straniero06} show that in low-metallicity stars
in the TP-AGB phase the abundance ratios of nuclei near the 
first and second \spro\ peaks are only weakly affected by the
efficiency of the $^{13}$C pocket (and hence the neutron source 
for the \spro), whereas abundances of species near the 
third \spro\ peak are strongly dependent on this parameter.
Therefore the abundance predictions of the \citet{arlandini99} stellar
model, which was computed for $\sim$~Solar metallicity stars 
in the TP-AGB phase, should predict the lighter \spro\ abundances 
at low metallicty better than it predicts the Pb abundance 
at low metallicity.
In a low-metallicity \spro\ nucleosynthesis event,
we would expect the $^{151}$Eu/Pb ratio to decrease relative to
a higher-metallicity case as a higher fraction of seed nuclei are 
converted to Pb.
Therefore, the stellar model pure-$s$ Eu/Pb and $^{151}$Eu/Pb ratios 
should represent upper limits for these values.
In both cases, decreasing these ratios would bring the stellar model
predictions into better agreement with the classical method predictions.

\section{Prospects for Future Studies of Neutron-Capture
Isotopic Fractions}
\label{future}

Previous studies have demonstrated that the Eu and Ba
isotopic fractions can be measured, 
we have shown that Sm isotopic fractions can 
also be quantitatively measured, and 
we also suggest that Nd isotopic fractions can 
be measured, albeit with lower precision that the other
rare-earth species.
The measurements of each set of isotopes give an
indication of the stellar nucleosynthetic history
that is consistent with other indicators, and
we now discuss the role that measurements of 
Ba, Nd, Sm, Eu, and other \ncap\ isotopic fractions
might play in future studies.

\subsection{Isotopic Fractions of Multiple Neutron-Capture Species}
\label{multipleiso}

It is a success of both the stellar model and classical method 
that these models' predictions of the
Sm $f_{152+154}$ isotopic fraction are in good agreement.
Suppose that both Sm isotopic fractions
$f_{152+154}$ and $f_{\rm odd}$ 
could be measured from the same set of lines.
This would enable a comparison
of the models' predicted abundances of
$^{147}$Sm, which differs by a factor of $\sim$\,3
and affects the predicted $f_{\rm odd}$ values,
whereas both models generally agree on
$^{149}$Sm, $^{152}$Sm, and $^{154}$Sm.
One could measure the isotopic
fractions of Sm and Ba in an $r$-enriched
star.
If a convincing argument could be made 
that the Ba and Sm should have originated
from the same nucleosynthetic event(s)
(i.e., an extra contribution has not been made to the
Ba abundances by another \ncap\ process),
the isotopic fractions measured
for both Ba and Sm should show the same 
relative proportion of \rpro\ isotopes.
This scenario could permit a direct test to the
predicted Ba isotopic fractions, which
differ by 26\% in their $r$-only
predictions for $f_{\rm odd}$,
with the most significant differences
arising in the models' predictions for
$^{137}$Ba and $^{138}$Ba.
Such tests would be greatly welcomed.

It is not obvious why Ba abundances
are sometimes larger than the scaled
S.~S.\ $r$-process abundances
(scaled to the heavier $n$-capture elements) 
in metal-poor stars with an
$r$-only chemical signature and no evidence of binarity,
such as HD~175305.  
(See also, e.g., Figure~15 of \citealt{honda04a}
or Figure~15 of \citealt{barklem05},
where some of their $r$-enriched stars
exhibit this behavior.)
\citet{asplund04} suggests that the 
3D Ba abundance correction may be 
as much as $-$0.5~dex for some metal-poor dwarfs
and subgiants.
While this matter may indeed be a result
of poor Ba abundance determinations
due to 1D LTE modeling or to uncertainties
in the S.~S.\ abundance predictions,
the search for an astrophysical explanation 
for this phenomenon should be explored further.
Measurements of Ba and other \ncap\ isotopic
fractions could be used to identify any 
correlations that may exist between 
super-$r$-process Ba abundances and
other nucleosynthetic signatures.

The combination of Ba, Nd, Sm, and Eu isotopic fractions
could provide more complete knowledge 
of abundances at the isotopic level
over a range of $\sim$~20 mass numbers
($A\,=\,$134$-$154).
This information would complement elemental abundance
measurements of lighter and heavier
$n$-capture species to get a more 
complete picture of the \ncap\ nucleosynthesis,
constrain the conditions (e.g., temperature, neutron density,
etc.) that are required to produce the \rpro\ elements, and
determine the actual \rpro\ path by identifying the 
individual isotopes that participate in this process
(see \citealt{cowan04} for further discussion).
Extended discussion of the relationship between neutron densities
in the ``main'' and ``weak'' $r$-processes
and the isotopic abundance signatures in this mass range 
is provided in \citet{kratz07} and will not be repeated here.
There is evidence that different mass regimes
may be synthesized by different astrophysical
processes in addition to the $s$- and $r$-processes,
such as the light element primary process, or LEPP
\citep{travaglio04}.
While the LEPP would have the greatest effect
on nucleosynthesis of lighter elements such as the
Sr-Y-Zr group, knowledge of isotopic data 
for elements with very different $Z$ would
still be extremely informative.

This study has made use of the most current and complete
laboratory atomic data to compute the isotope shifts
and HFS patterns for Sm and Nd.
While we are pleased that the available atomic data
permits measurement of the isotopic fractions
of these two species, complete sets of HFS constants 
and isotope shifts have not been measured for many lines 
that lie in the yellow-red regions of the spectrum.
These transitions are generally weaker in stellar spectra,
but the isotope shifts may be large enough nonetheless
to permit reliable measurements of the isotopic fraction
even with weak lines.
We encourage such efforts from the atomic physics community 
in the future.
We note that these lines may be significantly stronger
in more metal-rich stars or those which exhibit 
enhanced abundances for \spro\ elements, such as 
Ba stars \citep[e.g.,][]{allen06a,allen06b}, 
where the blue lines we have used in our
study are likely to be heavily blended with other
atomic and molecular features;
in these cases, lines in the yellow-red spectral
regions may provide the only opportunity to 
establish the \ncap\ signatures at the isotopic level.

\subsection{Individual Advantages and Disadvantages of 
Europium, Samarium, and Neodymium Isotopic Fractions}
\label{euvssm}

One advantage of using measurements of the Eu 
isotopic fraction 
as a nucleosynthesis probe
is the wide
HFS structure that can be measured with only moderately-high
spectral resolution ($R\,\sim$\,45--60,000).
In $r$-enriched stars, the Eu lines are strong
enough to facilitate measurement of the isotopic
fraction even at very low metallicities.
But, as we have argued in \S~\ref{euiso},
it is unlikely that Eu isotopes can be measured
with the precision to unambiguously distinguish
$s$- and \rpro\ nucleosynthesis using the
lines in the near-UV and blue regions of the spectrum
if indeed the \citet{arlandini99} predictions are correct.

Sm presents a different set of circumstances.
The HFS of Sm is not nearly as wide
as Eu, and the lines with the widest structure
are also very weak (or undetectable) 
and sometimes blended in very metal-poor stars.
Though we are forced to parametrize the
$s$- and \rpro\ nucleosynthetic predictions
into some set of isotopes (e.g., $f_{\rm odd}$ or
$f_{152+154}$) to facilitate distinguishing among seven isotopes
simultaneously, the greater difference 
predicted between $f^{s}_{152+154}$ and $f^{r}_{152+154}$ allows 
one to cleanly distinguish these different 
nucleosynthetic signatures with 
reasonable measurement uncertainties.
In addition, if Sm isotope fractions could be 
measured in a large sample of stars,
it may be possible to observe the changing
contributions of $s$- and \rpro\
nucleosynthesis in stars 
over the history of the galaxy 
(cf.\ Figure~6 of \citealt{mashonkina06}, 
who performed a similar study using Ba isotopes).
Therefore, while Sm lines that permit measurement of 
isotopic fractions may not be
accessible in all metal-poor stars,
we propose that in favorable cases Sm isotopes
may give a clearer picture of the 
\ncap\ nucleosynthetic history 
than Eu isotopes can.

The Nd isotopic fraction is more
difficult to measure reliably than the Sm isotopic fraction.
The smaller isotope shifts of Nd translate directly into
larger uncertainties in the measured isotopic fraction 
on any individual line.
This necessitates that as many clean lines as possible 
be measured in order to reduce the uncertainties
on the mean Nd isotopic fraction.
Only then is it possible to unambiguously 
distinguish distinct chemical signatures in the Nd isotopes.
Nd is produced more by the \spro\ than the \rpro,
though, so Nd lines may be easier to observe
in stars with an \spro\ chemical signature
than either Eu or Sm.

Of course these suggestions
only apply to the examination of the 
nucleosynthetic history at the isotopic level.
At the elemental level Eu should continue to
serve as an excellent barometer of 
\rpro\ nucleosynthesis, and Ba or La
should continue to serve as barometers of \spro\ nucleosynthesis.

\subsection{\spro\ Yields from Models of Low-Metallicity Stars on the AGB}
\label{lowzagb}

While the classical method of modeling the \spro\ is, 
by definition, model-independent, it has been shown
to predict an overproduction of $^{142}$Nd relative to 
other isotopes of Nd and pure-$s$ nuclei, 
e.g., as determined from studies
of the Murchison meteorite \citep{zinner91,guber97}.
This isotope is only produced by the \spro\ and 
contains 82 neutrons, one of the magic neutron numbers,
making this nucleus relatively more stable than other nuclei with
similar mass numbers and thus violating the classical method's 
assumption of a smoothly-varying $\sigma N_{s}$ curve.
The complex stellar model of \citet{arlandini99} can 
avoid this difficulty near the magic neutron numbers.

Of the much theoretical and computational work that has been
performed in recent years to better understand
the structure and evolution of low-metallicity stars on the AGB,
one goal has been to provide a complete set of chemical
yields from H to Bi, at the termination of the \spro.
The first such set of yields was presented by \citet{cristallo06},
for 2\,M$_{\sun}$ stars on the AGB with $Z=1.5\times10^{-2}$,
$1\times10^{-3}$, and $1\times10^{-4}$.
Similar models surely will follow, and the observational
verification for these models will rely heavily on isotopic abundances
as well as elemental abundances, particularly for magic neutron
number nuclei such as $^{142}$Nd.
The interpretation of \ncap\ nucleosynthetic signatures of the
first generations of stars will depend heavily on the 
reliability of these models.
As these models mature and are (hopefully) shown to be reliable predictors
of \spro\ nucleosynthesis at low metallicity, they will provide insight
into the structure and evolution of stars on the AGB that cannot be
obtained from the classical ad hoc approach to understanding the \spro.

\subsection{The Role of Future Studies of Neutron-Capture
Isotopic Fractions}
\label{bigtelescopes}

We suggest that the
features which are strongly blended with the \ncap\
lines in our stars 
(in particular, Ti, V, Fe, and Ni) may be 
somewhat diminished in very metal-poor, 
\ncap\ enhanced stars.
The requirements of high spectral resolution
and high S/N to clearly discern \ncap\
isotopic fractions are currently
difficult to achieve for the numbers of
such stars that have been detected by the 
HK Survey \citep{beers85,beers92} and the
Hamburg/ESO Survey \citep{wisotzki00,christlieb03},
owing to the faintness of these stars.
From a technical outlook,
observations of this nature will become more feasible
with high-resolution echelle spectrographs on 
the next generation of 20$-$30\,m telescopes.
Knowledge of the \ncap\ isotopic composition
of these stars would complement the information gleaned
from elemental abundances alone and further enhance
our understanding of stellar nucleosynthesis 
and chemical evolution throughout the early
history of our Galaxy.

\section{Conclusions}
\label{conclusions}

We have successfully measured the isotopic
fractions of Eu and Sm in HD~175305, and we have assessed 
the Nd isotopic fraction as well.
We find that the Sm isotopic fraction is suggestive of an
\rpro\ origin.
The Eu and Nd isotopic fractions are
unable to distinguish between an $r$- or \spro\ origin.
Along with measurements of elemental
abundances of \ncap\ species,
our Sm isotopic fraction reinforces
the assertion that HD~175305 has been 
enriched by \rpro\ nucleosynthesis.
Both the elemental and isotopic abundance distributions in HD~175305
join the growing preponderance of evidence 
that supports the hypothesis of a universal \rpro\ mechanism
for elements with $Z\geq56$.

We have measured the isotopic fraction of Sm in HD~196944, which
suggests an \spro\ origin.  
The Nd isotopic fraction is unable to distinguish
between the two processes.
The Sm isotopic fraction complements previous elemental 
abundance measurements and 
observed radial velocity variations of this star
to confirm that the enrichment pattern of HD~196944 is consistent with
transfer of \spro\ material from a companion star in the AGB phase of
stellar evolution.

We suggest that measurements of the Sm isotopic fraction will be more
advantageous than measurements of the Eu isotopic fraction alone.
The Sm isotopic fraction $f_{152+154}$ permits
a clearer distinction of pure-$s$- and pure-$r$-process
content than Eu isotopic fractions can.
There often are fewer blending features in the spectral
regions where the useful Sm lines are located,
and there are more Sm lines available for analysis.
Nd lines may offer more opportunities to measure 
an isotopic fraction in $s$-process-enriched
stars than either Eu or Sm, although it is extremely challenging
to discern the Nd isotopic fraction.
We propose that measurements
of isotopic fractions of multiple \ncap\ species 
in the same metal-poor star can
allow direct quantification of the 
relative $s$- and \rpro\ contributions
to \ncap\ material,
enable comparison of these contributions
as a function of increasing nuclear mass number,
test yields of low-metallicity stars on the AGB,
and provide tests of \spro\ nucleosynthesis
predictions.
We have conducted one such test for $^{151}$Eu using
Eu and Pb abundances collected from the literature,
finding that the \citet{arlandini99} stellar model
predictions are a better match than the classical method
predictions for $^{151}$Eu, although this conclusion
hinges on measurements of the Eu isotopic fraction in 
only two $(r+s)$-enriched stars.

Based on our results, we argue that the rare earth $s$- and \rpro\ 
abundance patterns, which are observed in metal-poor stars at the elemental
abundance level, are also present in the isotopic fractions.
This implies that our understanding of \ncap\ nucleosynthesis
is not wildly mistaken.
This result has often been tacitly assumed in studies of
nucleosynthesis and chemical abundance patterns in the 
Galactic halo, and we contend that increasing empirical evidence
for multiple rare earth species now exists to support
this hypothesis.

\acknowledgments

We are pleased to acknowledge the following individuals for their
encouraging and helpful discussions:
Carlos Allende Prieto,
Jacob Bean,
Anna Frebel,
David Lambert,
Martin Lundqvist,
Rob Robinson,
and Glenn Wahlgren.
We thank the referee for providing a number of insightful comments
on this manuscript.
I.~U.~R.\ wishes to thank the McDonald Observatory staff
for their hospitality and their
dedication to the observatory, its visitors,
and the West Texas community.
This research has made use of the 
NASA Astrophysics Data System (ADS),
NIST Atomic Spectra Database,
Vienna Atomic Line Database (VALD; \citealt{kupka99}), 
Two Micron All-Sky Survey (2MASS), 
and SIMBAD databases.
The reliability and accessibility of
these online databases is greatly appreciated. 
Funding for this project has been generously provided by 
the U.~S.\ National Science Foundation
(grant AST~03-07279 to J.~J.~C., 
grant AST~05-06324 to J.~E.~L., and 
grants AST~03-07495 and AST~06-07708 to C.~S.)
and by the Sigma Xi 
\textit{Grants-in-Aid-of-Research} program.

%% NOTE TO EDITOR: PLEASE INCLUDE THIS NEXT LINE IN THE 
%% MANUSCRIPT.  THANKS!
%% NEED AASTEX 5.2 TO INTERPRET THIS MACRO
%{\it Facilities:} \facility{Smith (2dCoud\'{e})}

\appendix

\section{Comments on Individual Transitions}
\label{comments}

Here we describe the individual characteristics of our
lines of interest as well as blending features.  
Where possible, we have tried to incorporate
the highest quality atomic data for these blending features,
including log($gf$) values calculated from
laser-induced fluorescence measurements
of the radiative level lifetimes and 
branching fractions measured from 
Fourier transform spectrometry.
We alert the reader 
that many of the studies published
before the early 1980s used less precise methods,
and we employ those values with some caution.

\subsection{Europium Lines}
\label{eulines}

The Eu~\textsc{ii} line at 4129\AA\ is 
a clean, relatively unblended line in 
HD~175305; this spectral region was not
observed in our other two stars.
The only significant blend is a Fe \textsc{i} line 
at 4129.4611\footnote{
For wavelengths of blended features,
we quote the number of digits to the right of the 
decimal point to the precision given in the reference
and the precision used in our syntheses.}\AA\ 
in the far blue wing;
the wavelength of this line was taken from 
\citet{nave94} and the log($gf$) value was taken
from \citet{fuhr05}, who scaled the value 
found by \citet{may74}.
We note that these values are not
critical to determining the isotopic fraction of Eu.

The Eu~\textsc{ii} line at 4205\AA\ is
strongly blended with CH (and, to a lesser extent,
CN) features, two V~\textsc{ii} lines, and 
a Y~\textsc{ii} line.
We adopt a CH linelist from the Kurucz database
and a CN linelist from the Plez database\footnote{
Available online:
\url{ftp://saphir.dstu.univ-montp2.fr/GRAAL/plez/CNdata}}.
Precise laboratory wavelengths and log($gf$) measurements
are not available for the V~\textsc{ii} line at 
4205.04\AA; we adopt the V~\textsc{ii} 4205.084\AA\
log($gf$) value from \citet{biemont89}.  
A log($gf$) value for the 
Y~\textsc{ii} line at 4204.694\AA\ can be calculated
from the oscillator strength presented in \citet{hannaford82}.
Performing a synthesis of this region of the Solar 
spectrum using an interpolated empirical
\citet{holweger74} model atmosphere,
we find rather poor agreement with the
\citet{kurucz84} Solar flux spectrum,
with the most significant discrepancies arising
from the CH bands and the wavelength of the 
Y~\textsc{ii} line.  
One would hope that in a very metal-poor star
with sub-solar C abundance (HD~122563) 
these effects would be greatly diminished;
the situation is much less hopeful for a C-enhanced
star (HD~196944).
Yet given the extent and severity of the blends of this
Eu line, as well as its weakness in HD~122563,
we are reluctant to trust any of the isotope fractions
derived from this line.
This is unfortunate because only this Eu line 
is strong enough to permit measurement of the
elemental abundance in our observed spectra
in HD~122563 and HD~196944.

The Eu~\textsc{ii} line at 4435\AA\ is 
strongly blended with a Ca~\textsc{i} line at 
4435.67\AA\ and mildly blended with a 
Ni~\textsc{i} line at 4435.33\AA.
In HD~122563 and HD~196944, this Eu line
is only a small sliver on the blue wing of the 
Ca line, and therefore no reasonable assessment of the 
elemental abundance
or isotopic fraction can be made.
In HD~175305 the Ca line is saturated, 
and we cannot produce a satisfactory fit
to the blended profile no matter what 
log($gf$) value we adopt for this line.
We adjust the log($gf$) value of the Ni
line to match the Solar spectrum,
yet an additional adjustment of $+$0.65 
dex to the Solar log($gf$) value 
is necessary to bring this line 
into rational agreement with our 
observed spectrum. 
Even then we are not satisfied with 
the fit of the blue wing of the Eu line.
Consequently, because such a small region of the spectrum
covered by the Eu line remains unblended,
we also disregard this line in our 
measurement of the isotopic fraction.

\subsection{Samarium Lines}
\label{smlines}

The Sm~\textsc{ii} line at 4424.34\AA\ is relatively strong
and is blended only slightly with a Cr~\textsc{i} line
at 4424.28\AA.
An unpublished log($gf$) value for this line, $-$0.63, was measured
by \citet{sobeck07}.
We measure Sm isotopic fractions from this line
in HD~175305 and HD~196944;
the line is too weak to measure well in HD~122563.

The Sm~\textsc{ii} line at 4467.34\AA\ is also
relatively strong and blended slightly with a weak
Fe~\textsc{i} line at 4467.43\AA\ and an unidentified
(in, e.g., the \citealt{moore66} atlas) line at 4467.21\AA.
No experimental log($gf$) value is available for the Fe line,
but we use an inverted Solar analysis to derive
a log($gf$) of $-$2.92 for this line.
The unidentified feature can be reasonably modeled as
an Fe line.  
The isotope shifts of this line are almost 
negligible and the HFS is also small, but 
\citet{ekeland75} used it in their analysis of
the Sm isotopic fraction in the Sun
This line is not covered in any of our stellar spectra, 
but we do use it to revisit the Sm isotopic fraction
in the Solar photosphere.

The Sm~\textsc{ii} 
4591.81\AA\ line lies near the edge of the blue wing of a 
strong Cr~\textsc{i} line at 4592.05\AA\ that is easily
accounted for in the synthesis.  
Of greater concern, however, is an apparently unidentified
blend at 4591.73\AA\ in HD~175305.
We do not measure an isotopic fraction from this line 
but suggest that it might be useful in future studies.
In HD~122563 this line just falls off an echelle order,
and in HD~196944 this line is too weak to observe.

The Sm~\textsc{ii} 4593.53\AA\ line 
is affected by a very weak 
Fe~\textsc{i} line at 4593.53\AA, whose log($gf$) value
is given by
\citet{fuhr05}, which has been scaled from \citet{may74}.
We also notice a small, unidentified blend in the 
blue wing of this line at 4593.42\AA;
a close examination of \citet{lundqvist07}'s 
spectrum of CS~31082-001 in Figure~5 shows a similar feature.
This Sm line is too weak to be observed in HD~122563, 
is observed but is too weak to assess the isotopic fraction
in HD~196944,
and, regrettably, in HD~175305 this line fell on a piece of the
CCD where we encountered unrecoverable 
flat-fielding errors during
the reduction process.

The Sm~\textsc{ii} 4595.28\AA\ line is severely blended with
a strong Fe~\textsc{i} line at 4595.3591\AA.
The log($gf$) value for this Fe line can be 
computed from the data in \citet{obrian91},
but we are hesitant to believe any Sm 
abundances derived from this line in any of our stars.

The Sm~\textsc{ii} line at 4604.17\AA\ is promising.
A Fe~\textsc{i} line at 4603.949\AA\
affects only the bluemost wing
of this line, and a weak Fe~\textsc{i} line blends with it
at 4604.24\AA; a log($gf$) value is available
for the latter of these two Fe lines 
in \citet{gurtovenko81} and the log($gf$) 
for the former is derived from an inverted Solar analysis
($-$3.15),
which provides a far better fit than the log($gf$)
rated an ``E'' in \citet{fuhr05}.
We measure an isotopic fraction for this line
in HD~175305, but the line is too weak in the other
two stars.

The Sm~\textsc{ii} 4693.63\AA\ line is blended with a 
Ti~\textsc{i} line at 4693.67\AA.  
Using our Ti~\textsc{i} abundance and 
the log($gf$) value for this line
from \citet{kuhne78}, we cannot produce a satisfactory
fit for the Sm line.
An additional unidentified blend occurs at 4693.75\AA\ 
in our spectrum of HD~175305.
Our spectra of HD~122563 and HD~196944 do not cover
this region.
We do not make a measurement of the 
elemental abundance or isotopic fraction
of Sm from this line in HD~175305, 
but we suggest that this line could be a candidate
for future Sm isotopic analyses.

\citet{lundqvist07} note that the Sm~\textsc{ii}
line at 4715.27\AA\ is blended with a Ti~\textsc{i} line
at 4715.230\AA.
A log($gf$) value is available in \citet{kuhne78}.
This line was not covered in any of our spectra, but
isotopic analysis of this Sm line may be possible.

The Sm~\textsc{ii} line at 4719.84\AA\ is covered in
all three of our spectra, but is only strong enough 
to measure in HD~175305 and HD~196944.
Full sets of hyperfine A and B constants are
not available for this line, so we compute the profile
using only the isotope shifts measured by \citet{lundqvist07}.
This line is blended with a weak La~\textsc{ii}
line at 4719.92\AA;
this La line was not included in the analysis of 
\citet{lawler01a}.
While the Sm line is much stronger than the La line
in both of our spectra, 
the uncertainty in the La log($gf$) value
and the lack of HFS structure information for the Sm line
lead us to disregard this line in our isotopic analysis.

The Sm~\textsc{ii} line at 5052.75\AA\ is not covered
in any of our spectra.
Full sets of hyperfine A and B constants are 
not available for this line, either, but we compute 
a synthesis using the isotope shifts only.  
This line is blended with a Ti~\textsc{i} line at
5052.87\AA\ and a Fe~\textsc{i} line at 5052.9814\AA.
A log($gf$) value is listed for the Ti line in the 
catalog of \citet{savanov90}, 
and a wavelength for the
Fe line is given in \citet{nave94}.
We suggest that these blends may be able
to be accounted for even without precise
atomic data, and the isotope shifts alone
could provide some indication
of the Sm isotopic fraction in metal-poor stars,
but measurement of an exact isotopic fraction is 
probably unwarranted here.

The weak Sm~\textsc{ii} line at 5069.47\AA\ 
is blended with a weak Fe~\textsc{i} line
at 5069.4233\AA, which is blended with a
weak Ti~\textsc{i} line at 5069.35\AA.
The wavelength for this Fe line is listed
in \citet{nave94}, but we cannot find an
experimental log($gf$) value for this line,
and the Ti blend prevents us from using
an inverted Solar analysis to derive a log($gf$)
value for either line.
We observe the Sm line in our spectra
of HD~122563 and HD~196944;
however, in both cases the Sm line is 
weak and the observed spectrum is littered
with small, undocumented blends in addition to
the Fe line noted here.
We recommend that any attempts to use this line
for Sm isotopic analysis proceed with great caution.

The Sm~\textsc{ii} line at 5103.09\AA\ is weak and
lies on the red edge of a strong Ni~\textsc{i} line
at 5102.964\AA.
A log($gf$) value for the Ni line
is given in \citet{doerr85}, and we have set the
wavelength of this line from analysis of the Solar spectrum.
Full sets of hyperfine A and B constants are not
available for this line, yet we compute a synthetic
spectrum using only the isotope shifts.
We can detect the Sm line in HD~175305,
but it is so overwhelmed by the Ni line that we 
cannot trust even our derived elemental abundance.
This Sm line is too weak to detect in HD~122563 
and HD~196944.

Our redmost Sm~\textsc{ii} line lies at 5104.48\AA.
It is blended by a Fe~\textsc{i} line at 5104.4375\AA.
\citet{nave94} quote a precise wavelength for this Fe line,
and its log($gf$) value is listed in \citet{fuhr05},
who scaled \citet{may74}'s value.
%$-$1.59.
When we synthesize this line in HD~175305,
we are forced to enhance the Sm abundance by
more than a factor of 2.
We have re-evaluated the log($gf$) value of the 
Fe blend by comparing with the Solar spectrum,
and find that we can reasonably only increase \citet{fuhr05}'s
log($gf$) value by $\approx$\,0.2 dex, which does little
to provide a better match between our observed
and synthetic spectra.
In addition, there is a significant unidentified
blend at 5104.59\AA\ in the red wing of our Sm line.
While it is clear that the Sm line is broadened by
HFS and isotopic shifts, we unfortunately are 
unable to measure an isotopic fraction.
This Sm line is too weak to detect in HD~122563 and
is not covered in our spectrum of HD~196944.

\subsection{Neodymium Lines}
\label{ndlines}

The Nd~\textsc{ii} line at 4177.32\AA\
is relatively strong, and only its far red wing is blended
with a strong Y~\textsc{ii} and Fe~\textsc{i} blend.
A log($gf$) value can be calculated for the 
Y~\textsc{ii} line at 4177.536\AA\ from the 
$f$-value in \citet{hannaford82}, and 
a log($gf$) value for the Fe~\textsc{i} line at
4177.5935\AA\ is given in \citet{obrian91}.
This blend is easily accounted for and has little effect
on the derived Nd isotopic fraction.
This Nd line was not covered in our spectra of
HD~122563 and HD~196944, but we measure
an isotopic fraction from this line in HD~175305.

The Nd~\textsc{ii} line at 4232.38\AA\ has 
two identified blends.
The Hf~\textsc{ii} line at 4232.386\AA\ lies
coincident with our Nd line; a log($gf$) value
for this line was measured by \citet{lundqvist06}.
A V~\textsc{i} line at 4232.46\AA\ also blends
with our Nd line, and a log($gf$) value %(of ``C'' rating)
is provided by \citet{martin88} in the NIST database.
The uncertainties in the abundances and transition
probabilities of these two lines, as well as 
their severity of blending with our Nd feature,
lead us to disregard this line in our spectrum
of HD~175305.
This line was not covered in our spectra of
HD~122563 and HD~196944.
The isotope shifts of this Nd line are noticeable, though,
so we report the hyperfine patterns for this line nonetheless.

The Nd~\textsc{ii} line at 4314.51\AA\ lies in a 
region where continuum placement is difficult due
to the presence of the CH G-band.  
We have not attempted to measure the Nd isotopic
fraction from this line in any of our stars,
although the Nd line is visible in HD~175305 and
HD~196944.
The isotope shifts of this line are rather large
($\sim$~0.035\AA), so we report the hyperfine patterns
for this line, in hopes that the Nd isotopic fraction
could be discerned from this line in a favorable star.
\citet{denhartog03} did not measure a log($gf$) value
for this line, but we derive log($gf$)$\,=\,-0.22$
from an inverted Solar analysis, assuming 
log$\,\varepsilon\,$(Nd)$_{\sun}\,=\,1.50$.

The Nd~\textsc{ii} line at 4358.16\AA\ is relatively strong
and unblended in HD~175305, and we use it to measure
an isotopic fraction.
This line was not covered in our spectra of 
HD~122563 and HD~196944.

The Nd~\textsc{ii} line at 4446.38\AA\ has the largest
isotope shifts of any of the Nd transitions for which
we report an isotopic fraction.
In HD~175305, this line is blended only with a weak
Gd~\textsc{ii} line at 4446.502\AA\ in the red wing,
and \citet{denhartog06} provide a log($gf$) value for this line.
In HD~196944, some unidentified blends are observed in both
the red and blue wings of this line, but it does not appear
that these blends affect our measurement of the 
isotope fraction.
We measure an isotopic fraction from this line in 
each of these two stars.
This line is observed in our high-S/N spectrum
of HD~122563, but it has a continuum depth of only
$\sim$~1.5\%, and we cannot measure an isotopic fraction
from this line.

The Nd~\textsc{ii} line at 4567.61\AA\ has a depth 
of only 4\% in HD~175305, and an isotopic fraction
can only marginally be deduced from this line.
We identify no blending features, although our
observed spectrum clearly shows some weaker
blends in the wings of this line.
This line was not covered in our observed spectra
of HD~122563 and HD~196944.

\section{Hyperfine Component Data for Samarium}
\label{smappendix}

We present hyperfine and isotopic components
for the 13 lines of Sm~\textsc{ii} described in 
\S~\ref{smlines} in Table~\ref{tab4-stub}.
Hyperfine A (magnetic dipole) and B 
(electric quadrupole) constants, which govern the
relative positions of the hyperfine components,
along with isotope shifts,
were taken first from \citet{masterman03} if available.
Constants for additional levels
were taken from
\citet{dorschel81},
\citet{young87}, and
\citet{villemoes95}.
We give preference to the 
hyperfine constants and isotope shifts determined
from radio frequency measurements or with a 
single-frequency laser, as in
\citet{masterman03} and \citet{beiersdorf95},
supplemented with isotope shifts measured
from FTS spectra in \citet{lundqvist07}
for lines not covered in earlier studies.

The relative strengths are calculated from the
LS (Russell-Saunders) angular momentum 
coupling formulae 
(e.g., \citealt[][p.\ 238]{condon53}),
where we have replaced the orbital angular momentum
quantum number ($L$) by the total electronic 
angular momentum ($J$),
the electron spin ($S$) by the nuclear spin ($I$),
and the total electronic angular momentum ($J$) by
the total atomic angular momentum ($F$).  
Thus, $F=I+J$, and the energy of the $J$ level
is split into a number of sublevels given by $F$,
which runs from $|I-J|$ to $I+J$.
The energy spacing between each component is proportional
to $F$.
The nucleus
may also have an electric quadrupole moment, 
which produces an additional shift (but not splitting)
in the energy levels.
The strongest relative components are those for which
$\Delta\, F\,=\,\Delta\,J$.
We normalize the strengths for each line such that
they sum to one.
Absolute transition wavenumbers are from \citet{lundqvist07}
if the lines were included in their study,
otherwise they are computed using the energy levels tabulated
by \citet{martin78}.
The center-of-gravity transition wavenumbers and air 
wavelengths are given in the first two columns, respectively.
Component wavenumbers are converted to air wavelengths
according to the formula given by \citet{edlen53}.

\section{Hyperfine Component Data for Neodymium}
\label{ndappendix}

We present hyperfine and isotopic components
for the 6 lines of Nd~\textsc{ii} described in 
\S~\ref{ndlines} in Table~\ref{tab5-stub}.
Hyperfine A and B constants and isotope shifts
were taken from \citet{rosner05}.
All calculations were performed analogously to 
the Sm case.
Center-of-gravity transition wavenumbers for the lines
of interest were taken from FTS measurements by 
\citet{blaise84}.

\clearpage

\begin{deluxetable}{lcccccc}
%\tabletypesize{\scriptsize}
%\rotate
\tablecaption{Comparison of Model Atmosphere Parameters
\label{tab1}}
\tablewidth{0pt}
\tablecolumns{7}
\tablehead{
\colhead{Reference}          &
\colhead{$T_{\rm eff}$}      &
\colhead{log($g$)}       &
\colhead{[M/H]}          &
\colhead{[Fe/H]}\tablenotemark{a} &
\colhead{$v_{\rm micro}$}    &
\colhead{$v_{\rm macro}$}    \\
\colhead{}               &
\colhead{(K)}            &
\colhead{}               &
\colhead{}               &
\colhead{}               &
\colhead{(\kmsec)} &   %\colhead{(km/s)} & 
\colhead{(\kmsec)}     %\colhead{(km/s)}   
}
\startdata
\multicolumn{7}{c}{HD 122563} \\
\hline
This study         & 4430 & 0.55 & $-$2.63 & $-$2.83 & 2.4  & 4.4     \\
\cite{fulbright00} & 4425 & 0.6  & \nodata & $-$2.6  & 2.75 & \nodata \\
\cite{johnson02a}  & 4450 & 0.50 & $-$2.65 & $-$2.75 & 2.30 & \nodata \\
\cite{honda04b}    & 4570 & 1.1  & \nodata & $-$2.77 & 2.2  & \nodata \\
\cite{simmerer04}  & 4572 & 1.36 & \nodata & $-$2.72 & 2.90 & \nodata \\
\hline
\multicolumn{7}{c}{HD 175305} \\
\hline
This study         & 4870 & 2.15 & $-$1.40 & $-$1.60 & 1.2  & 4.6     \\
\cite{fulbright00} & 4575 & 2.4  & \nodata & $-$1.30 & 1.4  & \nodata \\
\cite{burris00}    & 5100 & 2.50 & $-$1.50 & $-$1.40 & 1.40 & \nodata \\
\cite{cowan05}     & 5040 & 2.85 & \nodata & $-$1.48 & 2.00 & \nodata \\
\hline
\multicolumn{7}{c}{HD 196944} \\
\hline
This study        & 5170 & 1.80 & $-$2.26 & $-$2.46 & 1.7 & 6.8     \\
\cite{zacs98}     & 5250 & 1.7  & $-$2.0  & $-$2.45 & 1.9 & \nodata \\
\cite{aoki02}     & 5250 & 1.8  & \nodata & $-$2.25 & 1.7 & \nodata \\
\enddata
%\tablecomments{Put text here if necessary.}
\tablenotetext{a}{[Fe~\textsc{i}/H] is listed here.  We find 
[Fe~\textsc{ii}/H]~$=-$2.79 for HD~122563,
[Fe~\textsc{ii}/H]~$=-$1.62 for HD~175305, and
[Fe~\textsc{ii}/H]~$=-$2.43 for HD~196944.
}
\end{deluxetable}

\clearpage

\begin{deluxetable}{lcccccccc}

%\tabletypesize{\scriptsize}
%\rotate
\tablecaption{Isotope Mixes for Pure $s$- and $r$-process Material
and Measured Isotopic Fractions
\label{tab2}}
\tablewidth{0pt}
\tablecolumns{9}
\tablehead{
\colhead{} &
\multicolumn{2}{c}{stellar model\tablenotemark{a}} &
\colhead{} &
\multicolumn{2}{c}{classical method\tablenotemark{b}} &
\colhead{} &
\colhead{} &
\colhead{} \\
\cline{2-3} \cline{5-6}
\colhead{Species} &
\colhead{$s$-only} &
\colhead{$r$-only} &
\colhead{} &
\colhead{$s$-only} &
\colhead{$r$-only} &
\colhead{Solar System\tablenotemark{c}} &
\colhead{HD 175305} &
\colhead{HD 196944}
}
\startdata
Ba: $f_{\rm odd}\,=$   & 0.11 & 0.46 & & 0.09 & 0.72 & 0.178 & \nodata           & \nodata           \\ 
Nd: $f_{142+144}\,=$   & 0.67 & 0.32 & & 0.69 & 0.27 & 0.510 & 0.21$^{+0.56}_{-0.21}$ & 0.36$^{+0.64}_{-0.36}$ \\
Sm: $f_{152+154}\,=$ & 0.21 & 0.64 & & 0.20 & 0.64 & 0.495 & 0.51$\,\pm\,$0.08 & 0.35$\,\pm\,$0.14 \\ 
Eu: $f_{151}\,=$       & 0.54\tablenotemark{d} & 0.47 & & 0.04\tablenotemark{d} & 0.47 & 0.478 & 0.50$\,\pm\,$0.04 & \nodata \\
\enddata
%\tablecomments{Put text here if necessary.}
\tablenotetext{a}{\cite{arlandini99}}
\tablenotetext{b}{\cite{simmerer04,cowan06a}}
\tablenotetext{c}{\cite{iupac97}}
\tablenotetext{d}{The classical fit---based on assuming a 
smoothly-varying abundance curve---predicts a much smaller 
\spro\ contribution to $^{151}$Eu than does the stellar
\spro\ model of \citet{arlandini99}.  See \S~\ref{eupb} 
for further discussion.}
\end{deluxetable}

\clearpage

\begin{deluxetable}{ccccccccc}
%\tabletypesize{\scriptsize}
%\rotate
\tablecaption{Heavy $n$-capture Elemental Abundances in HD 175305
\label{tab3}}
\tablewidth{0pt}
\tablecolumns{9}
\tablehead{
\colhead{} &
\colhead{} &
\colhead{} &
\colhead{} &
\colhead{} &
\colhead{} &
\colhead{} &
\colhead{log($gf$)} &
\colhead{log $\varepsilon_{\sun}$} \\
\colhead{Species} &
\colhead{$Z$} &
\colhead{log $\varepsilon$} &
\colhead{[X/Fe]} &
\colhead{$\sigma$} &
\colhead{$N_{\rm lines}$} &
\colhead{log $\varepsilon_{\sun}$} &
\colhead{Ref.}  &
\colhead{Ref.}
}
\startdata
Ba \textsc{ii} & 56 & $+$0.91 & $+$0.40 & 0.16 & 1  & 2.13 & 1 & 10 \\
La \textsc{ii} & 57 & $-$0.29 & $+$0.20 & 0.13 & 3\tablenotemark{a} & 1.13 & 2 & 2 \\
Ce \textsc{ii} & 58 & $+$0.05 & $+$0.09 & 0.13 & 3  & 1.58 & 3 & 10 \\
Nd \textsc{ii} & 60 & $+$0.09 & $+$0.21 & 0.13 & 10 & 1.50 & 4 & 10 \\
Sm \textsc{ii} & 62 & $-$0.25 & $+$0.37 & 0.11 & 9  & 1.00 & 5 & 5 \\
Eu \textsc{ii} & 63 & $-$0.65 & $+$0.46 & 0.07 & 3  & 0.51 & 6 & 10 \\
Gd \textsc{ii} & 64 & $+$0.00 & $+$0.50 & 0.15 & 4  & 1.12 & 7 & 10 \\
Dy \textsc{ii} & 66 & $+$0.01 & $+$0.49 & 0.16 & 1  & 1.14 & 8 & 10 \\
Pb \textsc{i}  & 82 & $<+$0.10& $<+$0.45& \nodata &1& 1.95 & 9 & 10 \\
\enddata
\tablecomments{ 
Ionized species abundance ratios are referenced
to the Fe~\textsc{ii} abundance in Table~\ref{tab3}.
We adopt the 
Solar photospheric Fe abundance 
log\,$\varepsilon_{\sun}$~(Fe)~$=7.52$ 
from \cite{sneden91}.
}
\tablenotetext{a}{
We synthesize these La~\textsc{ii} lines 
using the HFS patterns listed in
Table~4 of \cite{ivans06}.
}
\tablerefs{
(1)~\citet{klose02};
(2)~\citet{lawler01a};
(3)~\citet{palmeri00};
(4)~\citet{denhartog03};
(5)~\citet{lawler06};
(6)~\citet{lawler01b};
(7)~\citet{denhartog06};
(8)~\citet{wickliffe00};
(9)~\citet{biemont00};
(10)~\citet{grevesse02}}
\end{deluxetable}

\clearpage

\begin{deluxetable}{cccccc}
%\tabletypesize{\scriptsize}
\tablecaption{Isotopic Hyperfine Structure Patterns for Samarium
\label{tab4-stub}}
\tablewidth{0pt}
\tablecolumns{6}
\tablehead{
\colhead{Wavenumber} &
\colhead{Wavelength} &
\colhead{Component Position} &
\colhead{Component Position} &
\colhead{Isotope} &
\colhead{Strength\tablenotemark{a}} \\
\colhead{(cm$^{-1}$)} &
\colhead{(\AA)} &
\colhead{(cm$^{-1}$)} &
\colhead{(\AA)} &
\colhead{} &
\colhead{}
}
\startdata
 22595.910 & 4424.3373 &  0.20775 & -0.04067 & 147 & 0.028106 \\
 22595.910 & 4424.3373 &  0.13532 & -0.02649 & 147 & 0.001521 \\
 22595.910 & 4424.3373 &  0.06911 & -0.01353 & 147 & 0.000040 \\
 22595.910 & 4424.3373 &  0.15704 & -0.03074 & 147 & 0.023908 \\
 22595.910 & 4424.3373 &  0.09083 & -0.01778 & 147 & 0.002540 \\
 22595.910 & 4424.3373 &  0.03117 & -0.00610 & 147 & 0.000097 \\
 \vdots    & \vdots    &  \vdots  & \vdots   & \vdots & \vdots \\
\enddata
\tablecomments{
The full table is available in machine-readable form in the electronic
edition of the journal;
only a small portion is shown here to present
its general form and content.
For lines without full sets of hyperfine
A and B constants available
(4719, 5052, and 5103), 
for the $^{147}$Sm and $^{149}$Sm isotopes
we list only the isotope shifts, computed
as if all A and B values are zero.
The isotope shift for the $^{144}$Sm isotope of
the 5069 line could not be measured in the 
FTS spectrum of \cite{lundqvist07}. 
We estimate this value by scaling the
energy shifts of the $^{144}$Sm isotope 
relative to the heavier isotopes from
the 5052 and 5103 lines, which 
show similar structure patterns
(M.\ Lundqvist, 2007, private communication).}
\tablenotetext{a}{
The sum of the strength values for each 
line have been normalized to one.  
When using these components in our spectral synthesis,
we divide out the Solar System isotopic ratios 
as given by \cite{iupac97}, but we stress that
this step is NOT included in the data presented here.
}
\end{deluxetable}

\clearpage

\begin{deluxetable}{cccccc}
%\tabletypesize{\scriptsize}
\tablecaption{Isotopic Hyperfine Structure Patterns for Neodymium
\label{tab5-stub}}
\tablewidth{0pt}
\tablecolumns{6}
\tablehead{
\colhead{Wavenumber} &
\colhead{Wavelength} &
\colhead{Component Position} &
\colhead{Component Position} &
\colhead{Isotope} &
\colhead{Strength\tablenotemark{a}} \\
\colhead{(cm$^{-1}$)} &
\colhead{(\AA)} &
\colhead{(cm$^{-1}$)} &
\colhead{(\AA)} &
\colhead{} &
\colhead{}
}
\startdata
 23932.050 & 4177.3196 &  0.10001 & -0.01746 & 143 & 0.024146 \\
 23932.050 & 4177.3196 &  0.14184 & -0.02476 & 143 & 0.001719 \\
 23932.050 & 4177.3196 &  0.05369 & -0.00937 & 143 & 0.019886 \\
 23932.050 & 4177.3196 &  0.17828 & -0.03112 & 143 & 0.000061 \\
 23932.050 & 4177.3196 &  0.09013 & -0.01573 & 143 & 0.002841 \\
 23932.050 & 4177.3196 &  0.01208 & -0.00211 & 143 & 0.016161 \\
 \vdots    & \vdots    &  \vdots  & \vdots   & \vdots & \vdots \\
\enddata
\tablecomments{
The full table is available in machine-readable form in the electronic
edition of the journal;
only a small portion is shown here to present
its general form and content.
}
\tablenotetext{a}{
The sum of the strength values for each 
line have been normalized to one.  
When using these components in our spectral synthesis,
we divide out the Solar System isotopic ratios 
as given by \cite{iupac97}, but we stress that
this step is NOT included in the data presented here.
}
\end{deluxetable}

\clearpage

\clearpage

\clearpage

\clearpage

\clearpage

\clearpage

\clearpage

\clearpage

\clearpage

\clearpage

\clearpage

\clearpage

\clearpage

\clearpage

\clearpage

\clearpage

\end{document}